\pdfoutput=1

\documentclass[11pt,twoside,a4paper,cmspaper,final,collab]{cms-tdr}
\def\svnVersion{473814P}\def\cmsCernNoTag{CERN-EP-2018-126}\def\cmsCernDate{\today}\def\cmsMessage{Published in the European Physical Journal C as \href{http://dx.doi.org/10.1140/epjc/s10052-018-6148-7}{\doi{10.1140/epjc/s10052-018-6148-7.}}}

\begin{document}\cmsNoteHeader{SMP-16-007}

\hyphenation{had-ron-i-za-tion}
\hyphenation{cal-or-i-me-ter}
\hyphenation{de-vices}
\hyphenation{cal-or-i-me-ter}
\hyphenation{de-vices}
\RCS$Revision: 470959 $
\RCS$HeadURL: svn+ssh://svn.cern.ch/reps/tdr2/papers/SMP-16-007/trunk/SMP-16-007.tex $
\RCS$Id: SMP-16-007.tex 470959 2018-08-05 14:32:30Z alverson $
\newlength\cmsFigWidth
\ifthenelse{\boolean{cms@external}}{\setlength\cmsFigWidth{0.42\textwidth}}{\setlength\cmsFigWidth{0.48\textwidth}}
\newlength\cmsFigWidthBig
\ifthenelse{\boolean{cms@external}}{\setlength\cmsFigWidthBig{0.6\textwidth}}{\setlength\cmsFigWidthBig{0.7\textwidth}}
\ifthenelse{\boolean{cms@external}}{\providecommand{\cmsLeft}{top\xspace}}{\providecommand{\cmsLeft}{left\xspace}}
\ifthenelse{\boolean{cms@external}}{\providecommand{\cmsRight}{bottom\xspace}}{\providecommand{\cmsRight}{right\xspace}}

\newlength\cmsTabSkip\setlength{\cmsTabSkip}{1ex}

\newcommand{\zmumu}{\ensuremath{\cPZ/\gamma\to \MM}\xspace}
\newcommand{\zee}{\ensuremath{\cPZ/\gamma\to \EE}\xspace}
\newcommand{\ztautau}{\ensuremath{\cPZ/\gamma\to \TT}\xspace}
\newcommand{\zll}{\ensuremath{\cPZ/\gamma\to \ell^+\ell^-}\xspace}
\newcommand{\csh}{\ensuremath{\cos\theta^{*}}\xspace}
\newcommand{\mll}{\ensuremath{m_{\ell\ell}}\xspace}
\newcommand{\yll}{\ensuremath{y_{\ell\ell}}\xspace}
\newcommand{\ptll}{\ensuremath{p_{\mathrm{T},\ell\ell}}\xspace}
\newcommand{\pzll}{\ensuremath{p_{z,\ell\ell}}\xspace}
\newcommand{\sineff}{\ensuremath{\sin^2\theta^{\ell}_{\text{eff}}}\xspace}
\newcommand{\absyll}{\ensuremath{\abs{y_{\ell\ell}}}\xspace}
\newcommand{\abseta}{\ensuremath{\abs{\eta}}\xspace}

\newcommand{\txtF}{\ensuremath{\mathrm{F}}\xspace}
\newcommand{\txtB}{\ensuremath{\mathrm{B}}\xspace}
\newcommand{\txtf}{\ensuremath{\mathrm{f}}\xspace}
\newcommand{\txtD}{\ensuremath{\mathrm{D}}\xspace}
\newcommand{\txtN}{\ensuremath{\mathrm{N}}\xspace}
\newcommand{\txtR}{\ensuremath{\mathrm{R}}\xspace}
\newcommand{\txtmin}{\ensuremath{\text{min}}\xspace}
\newcommand{\txteff}{\ensuremath{\text{eff}}\xspace}

\cmsNoteHeader{SMP-16-007}
\title{Measurement of the weak mixing angle using the forward-backward asymmetry of Drell--Yan events in $\Pp\Pp$ collisions at 8\TeV}
\titlerunning{Measurement of the weak mixing angle using the forward-backward asymmetry of Drell--Yan events}

\abstract{
A measurement is presented of the effective leptonic weak mixing angle (\sineff) using the forward-backward asymmetry of Drell--Yan lepton pairs ($\mu\mu$ and \Pe\Pe) produced in proton-proton collisions at $\sqrt{s}=8\TeV$ at the CMS experiment of the LHC. The data correspond to integrated luminosities of 18.8 and $19.6\fbinv$ in the dimuon and dielectron channels, respectively, containing 8.2 million dimuon and 4.9 million dielectron events. With more events and new analysis techniques, including constraints obtained on the parton distribution functions from the measured forward-backward asymmetry, the statistical and systematic uncertainties are significantly reduced relative to previous CMS measurements. The extracted value of \sineff from the combined dilepton data is
$\sineff=0.23101\pm0.00036\stat\pm0.00018\syst\pm0.00016\thy\pm0.00031\,\text{(parton distributions in proton)}=0.23101 \pm 0.00053$.
}

\hypersetup{
pdfauthor={CMS Collaboration},
pdftitle={Measurement of the weak mixing angle with the forward-backward asymmetry of Drell-Yan events at 8 TeV},
pdfsubject={CMS},
pdfkeywords={CMS, physics, electroweak, mixing angle}}

\maketitle

\section{Introduction \label{introduction}}

We report a measurement of the effective leptonic weak mixing angle (\sineff) using the forward-backward asymmetry (\AFB) in Drell--Yan
$\qqbar\to\ell^+\ell^-$ events, where $\ell$ stands for muon ($\mu$) or electron (\Pe). The analysis is based on data from the CMS experiment at the CERN LHC. At leading order (LO), lepton pairs are produced through the annihilation of a quark with its antiquark into a \PZ boson or a virtual photon: $\qqbar\to \cPZ/\gamma\to \ell^+\ell^-$. For a given dilepton invariant mass $m_{\ell\ell}$, the differential cross section at LO can be expressed at the parton level as
\begin{equation}
	\label{eq:crosssection}
	\frac{\rd\sigma}{\rd(\cos\theta^{*})} \propto 1 + \cos^{2} \theta^{*} + A_{4}  \cos\theta^{*},
\end{equation}
where the $(1 + \cos^{2} \theta^{*})$ term  arises from the spin-1 of the exchanged boson, and the $\cos\theta^{*}$ term originates from interference between vector and axial-vector contributions.
The definition of \AFB is based on the angle $\theta^*$ of the negative lepton ($\ell^-$) in the Collins--Soper~\cite{Collins} frame of the dilepton system:
\begin{equation}
    \label{eq:afb}
    \AFB=\frac{3}{8}A_4  =\frac{\sigma_\txtF-\sigma_\txtB}{\sigma_\txtF+\sigma_\txtB},
\end{equation}
where $\sigma_\txtF$ and $\sigma_\txtB$ are, respectively, the cross sections in the forward ($\csh>0$) and backward ($\csh<0$) hemispheres.
In this frame, $\theta^*$ is the angle of the $\ell^-$ relative to the axis that bisects the angle between the direction of the quark and the reversed direction of the antiquark. In proton-proton ($\Pp\Pp$) collisions, the direction of the quark is more likely to be in the direction of the Lorentz boost of the dilepton. Therefore, \csh can be calculated using the following variables in the laboratory frame:
\begin{equation}
    \csh=\frac{2(P_1^+P_2^- - P_1^-P_2^+)}{\sqrt{\mll^2(\mll^2+\ptll^2)}}\,\frac{\pzll}{\abs{\pzll}},
\end{equation}
where \mll, \ptll, and \pzll are the mass, transverse momentum, and longitudinal momentum, respectively, of the dilepton system, and the $P_i^\pm$ are defined in terms of the energies~($E_i$) and longitudinal momenta~($p_{z, i}$), of the negatively  and positively charged leptons as $P_i^\pm=(E_i\pm p_{z,i})/\sqrt{2}$~\cite{Collins}.

A non-zero \AFB value in dilepton events arises from the vector and axial-vector couplings of electroweak bosons to fermions.
At LO, these respective couplings of \PZ bosons to fermions (f) can be expressed as:
\begin{gather}
    \begin{align}
	v_\txtf&= T_3^\txtf-2Q_\txtf\sin^2\theta_\PW, \\
	a_\txtf&= T_3^\txtf,
    \end{align}
\end{gather}
where $Q_\txtf$ and $T_3^\txtf$ are the charge and the third component of the weak isospin of the fermion, respectively, and $\sin^2\theta_\PW$ refers to the weak mixing angle, which is related to the masses of the \PW\ and \PZ bosons through the relation $\swsq=1-m_\PW^2/m_\PZ^2$.
Electroweak (EW) radiative corrections affect these LO relations.
In the improved Born approximation~\cite{Bardin,Zfitter}, some of the higher-order corrections are absorbed into an effective mixing angle.
The effective weak mixing angle is based on the relation
$v_\txtf/a_\txtf=1-4\abs{Q_\txtf}\sin^2\theta_{\txteff}^\txtf$,
with
$\sin^2\theta_\txteff^\txtf=\kappa_\txtf \sin^2\theta_\PW$,
where the flavor-dependent $\kappa_\txtf$ is determined through EW corrections.
The \AFB for dilepton events is sensitive primarily to \sineff.

We measure \sineff by fitting the mass and rapidity $(\yll)$ dependence of the observed \AFB in dilepton events to standard model (SM) predictions as a function of \sineff. The most precise previous measurements of \sineff were performed by the combined LEP and SLD experiments~\cite{ALEPH:2005ab}. There is, however, a known discrepancy of about 3 standard deviations between the two most precise values. Other measurements of \sineff have also been reported by the Tevatron and LHC experiments~\cite{Abazov:2008xq,Abazov:2011ws,Chatrchyan:2011ya,Aaltonen:2013wcp,Aaltonen:2014loa,Abazov:2014jti,Aad:2015uau,Aaij:2015lka,Aaltonen:2016nuy,Abazov:2017gpw,Aaltonen:2018dxj}.

Using the LO expressions for the \PZ boson, virtual photon exchange, and their interference, the ``true'' $\AFB$ (\ie, using the quark direction in the definition of \csh) can be evaluated as
\ifthenelse{\boolean{cms@external}}{
\begin{multline} \label{eq:trueafb}
    \AFB^{\text{true}}(\mll)= a_\ell a_\cPq (8v_\ell v_\cPq - Q_\cPq KD_m)\\
    \times [16(v_\ell^2+a_\ell^2)(v_\cPq^2+a_\cPq^2)-8 v_\ell v_\cPq Q_\cPq KD_m\\
    + Q_\cPq^2K^2(D_m^2+\Gamma^2_\PZ/m_\PZ^2)]^{-1},
\end{multline}
}{
\begin{equation} \label{eq:trueafb}
    \AFB^{\text{true}}(\mll)=\frac{6 a_\ell a_\cPq (8v_\ell v_\cPq - Q_\cPq KD_m)}{16(v_\ell^2+a_\ell^2)(v_\cPq^2+a_\cPq^2)-8 v_\ell v_\cPq Q_\cPq KD_m+ Q_\cPq^2K^2(D_m^2+\Gamma^2_\PZ/m_\PZ^2)},
\end{equation}
}
where the subscript \cPq\ refers to the participating quark, $K=8\sqrt{2}\pi \alpha/G_\txtF m_\PZ^2$, $D_m=1-m_\PZ^2/\mll^2$, $\alpha$ is the electromagnetic coupling, $G_\txtF$ is the Fermi constant, and $\Gamma_\PZ$ is the full decay width of the \PZ boson.
A strong dependence of \AFB on \mll originates from axial and vector interference.
The \AFB is negative at small \mll and positive at large values, crossing $\AFB=0$ slightly below the \PZ boson peak.

\begin{figure*}[!htbp]
\centering
    \includegraphics[width=0.32\textwidth]{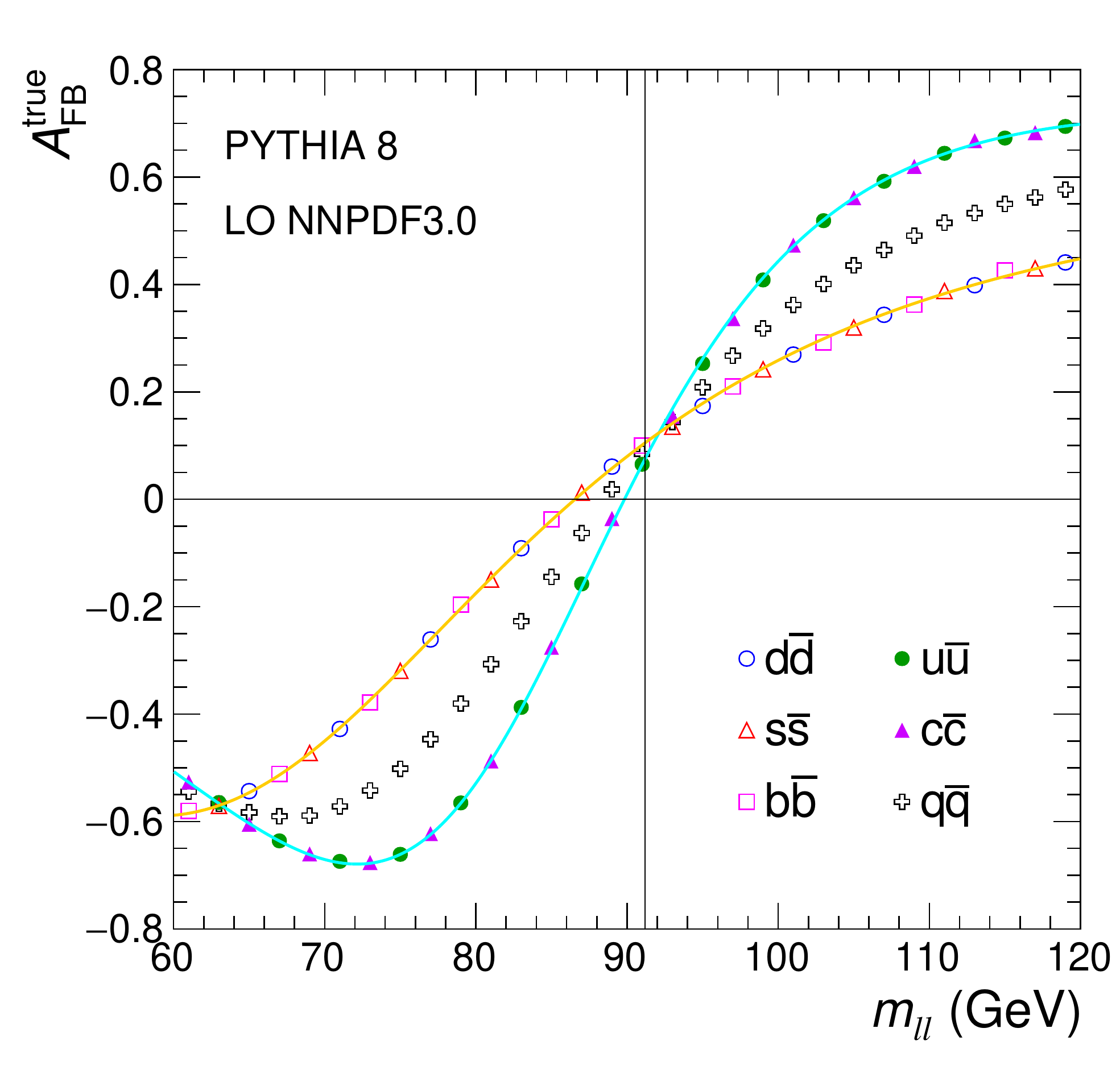}
    \includegraphics[width=0.32\textwidth]{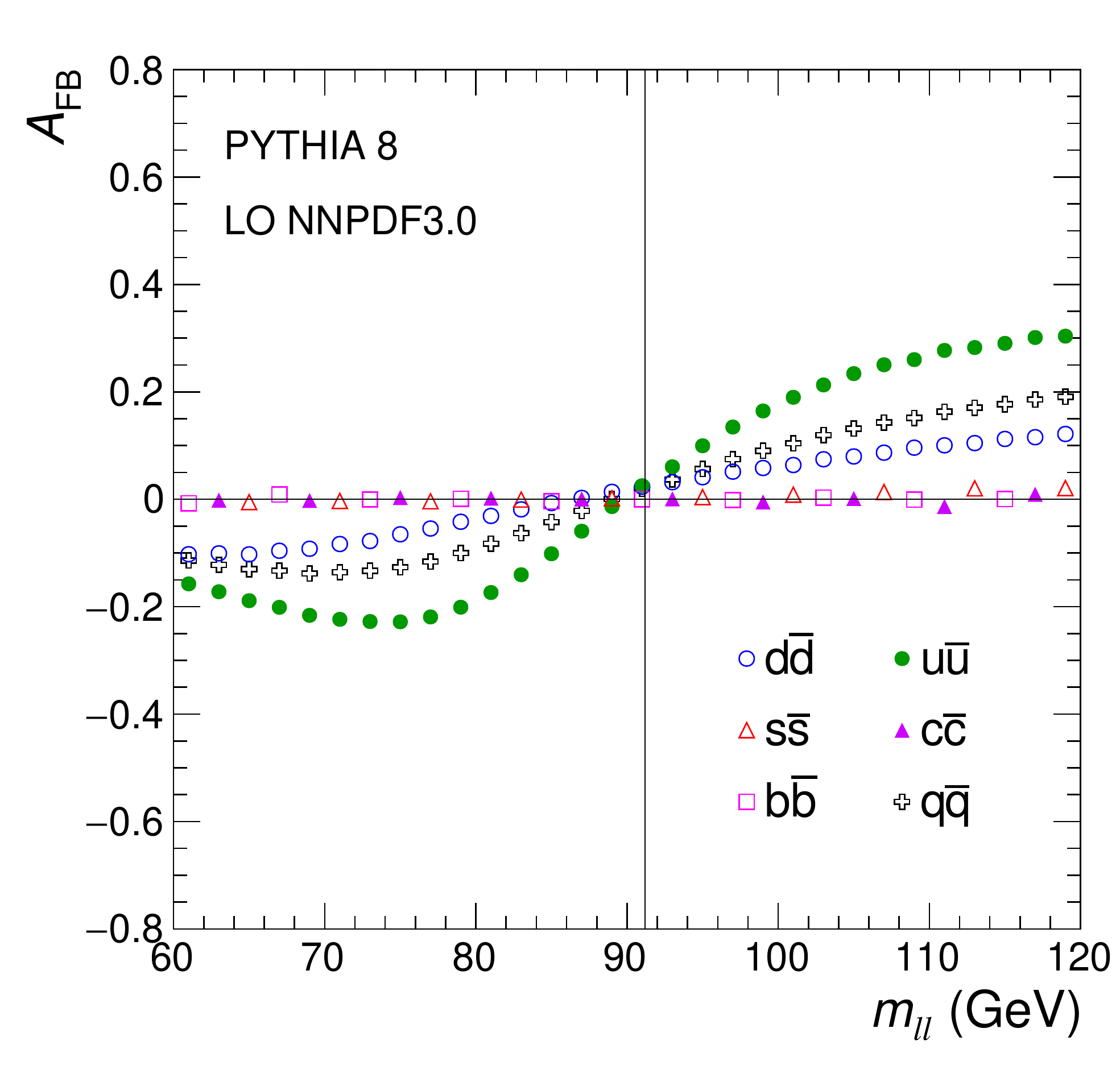}
    \includegraphics[width=0.32\textwidth]{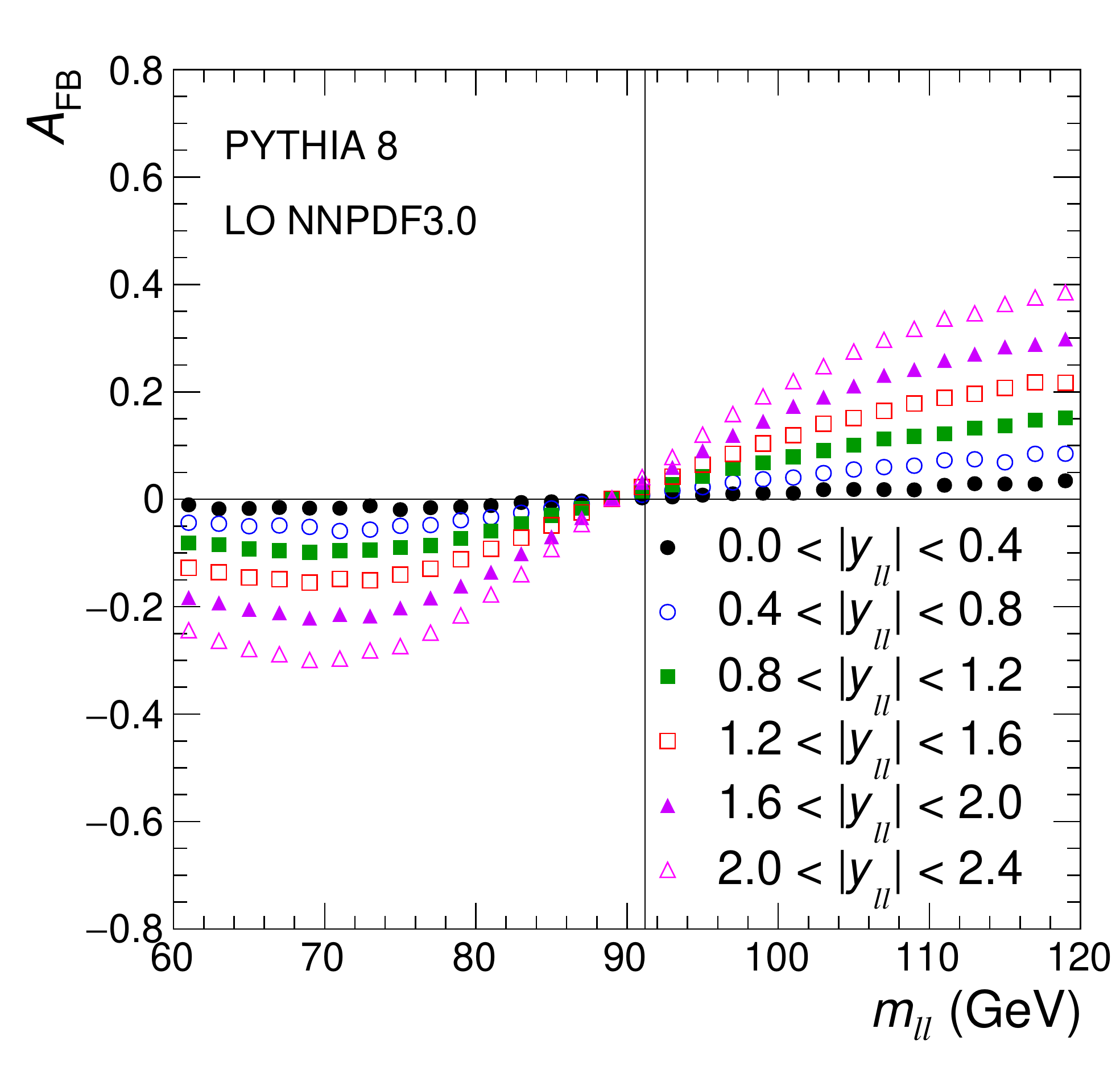}
    \caption{
	The dependence of $\AFB$ on $\mll$ in dimuon events generated using \PYTHIA~8.212~\cite{PYTHIA8}
	and the LO NNPDF3.0~\cite{NNPDF30} PDFs for dimuon rapidities of $\absyll<2.4$.
	The distributions for the total production (\qqbar) and the different channels are given on the left,
	overlaid with results based on Eq.~(\ref{eq:trueafb}), using the definition of $\AFB^\text{true}(\mll)$ for the known quark direction.
	The middle panel gives the diluted \AFB using instead the direction of the dilepton boost,
	and the right panel shows the diluted \AFB in $\absyll$ bins of 0.4 for all channels.
	\label{figure:typicalAFB}
    }
\end{figure*}

In collisions of hadrons, \AFB is sensitive to parton distribution functions (PDFs) for two reasons. First, the different couplings of \cPqu- and \cPqd-type quarks to EW bosons generate different \AFB values in the corresponding production channels, which means that the average depends on the relative contributions of \cPqu- and \cPqd-type quarks to the total cross section. Second, the definition of \AFB in $\Pp\Pp$ collisions is based on the sign of \yll, which relies on the fact that on average the dilepton pairs are Lorentz-boosted in the quark direction. Therefore, a non-zero average \AFB originates only from valence-quark production channels and is diluted by events where the antiquark carries a larger momentum than the quark. A dependence of the ``true'' and diluted \AFB on dilepton mass for different \qqbar production channels and their sum is shown in Fig.~\ref{figure:typicalAFB}.

The dilution of \AFB depends strongly on \yll, as shown in Fig.~\ref{figure:typicalAFB}. At zero rapidity, the quark and antiquark carry equal momenta, and the dilution is maximal, resulting in $\AFB=0$. The \AFB is measured in 12 bins of dilepton mass, covering the range $60<\mll<120\GeV$, and 6 $\absyll$ bins of equal size for $\absyll<2.4$. The boundaries in the dilepton mass are at: 60, 70, 78, 84, 87, 89, 91, 93, 95, 98, 104, 112, and 120\GeV. The mass bins are chosen such that near $m_\PZ$ the bin widths are larger than the mass resolution in any of the ranges of \yll. Smaller and larger mass bins are chosen such that all mass bins contain enough events to perform a meaningful independent measurement. The weak dependence of \AFB on \ptll is included in the SM predictions. The uncertainty originating from modeling of \ptll is very small and included in the theoretical estimates.

\section{The CMS detector}
The central feature of the CMS apparatus is a superconducting solenoid of 6\unit{m} internal diameter, providing a magnetic field of 3.8\unit{T}.
A silicon pixel and strip tracker, a lead tungstate crystal electromagnetic calorimeter (ECAL), and a brass and scintillator hadron calorimeter (HCAL), each composed of a barrel and two endcap sections reside within the solenoid volume. Forward calorimeters extend the pseudorapidity $\eta$ coverage provided by the barrel and endcap detectors. Muons are measured in gas-ionization detectors embedded in the steel flux-return yoke outside the solenoid. A more detailed description of the CMS detector can be found in Ref.~\cite{Chatrchyan:2008zzk}.

Muons are measured in the range $\abs{\eta} < 2.4$, using detection planes based on the drift-tube, cathode-strip chamber, or resistive-plate chamber technologies. Matching muons to tracks measured in the silicon tracker provides a relative transverse momentum resolution for muons with $20 <\pt < 100\GeV$ of 1.3--2.0\% in the barrel, and less than 6\% in the endcaps. The \pt resolution in the barrel is smaller than 10\% for muons with \pt up to 1\TeV~\cite{Chatrchyan:2012xi}.

The electromagnetic calorimeter consists of 75\,848 lead tungstate crystals that provide a coverage of $\abs{\eta} < 1.48 $ in the barrel region and $1.48 < \abs{\eta} < 3.00$ in the two endcap regions. Preshower detectors consisting of two planes of silicon sensors, interleaved with a total of 3~radiation lengths of lead, are located in front of each endcap detector. The electron momentum is obtained by combining the energy measurement in the ECAL with that in the tracker. The momentum resolution for electrons with $\pt \approx 45\GeV$ from $\Z \to \Pe \Pe$ decays, ranges from 1.7\% for nonshowering electrons in the barrel region, to 4.5\% for showering electrons in the endcaps~\cite{Khachatryan:2015hwa}.

Events of interest are selected using a two-tiered trigger system~\cite{Khachatryan:2016bia}. The first level, consisting of custom hardware processors, uses information from the calorimeters and muon detectors to select events at a rate of about 100\unit{kHz} within a time interval of less than 4\mus. The second level, known as the high-level trigger, consists of a farm of processors running a version of the full event reconstruction software optimized for fast processing, that reduces the event rate to about 1\unit{kHz} before data storage.

\section{Data and simulated events}
The measurement is based on $\Pp\Pp$ collisions at $\sqrt{s}=8\TeV$ recorded by the CMS Experiment in 2012, corresponding to integrated luminosities of 18.8 and 19.6\fbinv for muon and electron channels, respectively.

Candidates for the dimuon channel are collected using an isolated single-muon trigger with a \pt threshold of 24\GeV and $\abseta<2.4$. At the beginning of data taking, the muon trigger was restricted to $\abseta<2.1$. We do not use these events, and the integrated luminosity in the dimuon analysis is therefore somewhat smaller than for dielectrons. Background contamination is reduced by applying identification and isolation criteria to the reconstructed muons. First, muon tracks are required to be reconstructed independently in the inner tracker and in the outer muon detectors. A global fit to the momentum, including both tracker and muon detector hits, must have a fitted $\chi^2/\text{dof}<10$, where dof stands for the degrees of freedom. Muon tracks are required to pass within a transverse distance of $0.2\cm$ from the primary vertex, defined as the $\Pp\Pp$ vertex with the largest $\sum \pt^2$ of its associated tracks. Muon candidates are rejected if the scalar-\pt sum of all tracks within a cone of $\Delta R=\sqrt{\smash[b]{(\Delta\eta)^2+(\Delta\phi)^2}}=0.3$ around the muon is larger than 10\% of the \pt of the muon (this is referred to as track isolation, with $\phi$ being the azimuth in radians). The track isolation requirement  is insensitive to contributions from additional soft $\Pp\Pp$ interactions (pileup). An event is selected when there are at least two isolated muons, with the leading muon (\ie, the one with largest \pt) having $\pt>25\GeV$, and the next-to-leading muon having $\pt>15\GeV$. At least one muon with $\pt>25\GeV$ is required to trigger the event. For the Drell--Yan signal, the two leptons are required to have opposite sign (OS).

{\tolerance=1200
Dielectron candidates are collected using a single-electron trigger with a \pt threshold of 27\GeV and $\abseta<2.5$.
Variables pertaining to the energy distribution in electromagnetic showers and to impact parameters of inner tracks are used to separate prompt electrons from electrons originating from photon conversions in detector material.
The jet background from SM events produced through quantum chromodynamics (QCD) is referred to as multijet production.
A particle-flow (PF) event reconstruction algorithm is used to identify different particle types (photons, electrons, muons, and charged and neutral hadrons~\cite{CMS-PRF-14-001}). The scalar-\pt sum of all PF particles in a cone of $\Delta R<0.3$ around the electron direction is required to be less than 15\% of the electron \pt, which reduces the background from hadrons in multijet events that are reconstructed incorrectly as electrons. This sum is corrected for contributions from pileup~\cite{Khachatryan:2015hwa}. The electron momentum is evaluated by combining the energy in the ECAL with the momentum in the tracker. To ensure good reconstruction, the coverage is restricted to $\abseta<2.4$, excluding the transition region of $1.44<\abseta<1.57$ between the ECAL barrel and endcap detectors, as electron reconstruction in this region is not optimal. Dielectron candidates are selected when at least two OS electrons pass all quality requirements. The leading and next-to-leading electrons must have respectively $\pt>30$ and  $>20\GeV$, with the triggering electron always required to have $\pt>30\GeV$.
\par}

A total of about 8.2 million dimuon and 4.9 million dielectron candidate events are selected for further analysis. The number of dielectron events is smaller because of the higher \pt thresholds and more stringent selection criteria implemented in electron selections.
The \zmumu and \zee data include small (${<}1\%$) background contaminations that originate from \ztautau, \ttbar, single top quark, and diboson (\PW\PW, \PW\cPZ, and \cPZ\cPZ) events, as well as multijet and \PW$+$jets events. Contributions from these backgrounds are subtracted from data as described below. Contamination from photon-induced background near the \PZ boson peak is negligible~\cite{Bourilkov:2016qum}.

Monte Carlo (MC) simulation is used to model signal and background processes. The signal as well as the single-boson and top quark backgrounds are based on next-to-leading order (NLO) matrix elements implemented in the \POWHEG~v1 event generator~\cite{POWHEG0,POWHEG1,POWHEG2,POWHEG3} using the CT10~\cite{CTEQ:1007} PDFs. The generator is interfaced to \PYTHIA~6.426~\cite{PYTHIA6} using the Z2*~\cite{Chatrchyan:2013gfi,Khachatryan:2015pea} underlying event tune, which generates the parton showering, the hadronization, and the electromagnetic final-state radiation (FSR). The background events from $\tau$ lepton decays are simulated with \TAUOLA 2.7~\cite{TAUOLA}. Diboson and multijet background events are generated with \PYTHIA 6 using the CTEQ6L1 PDFs~\cite{CTEQ6L}.
Simulated minimum-bias events are superimposed on the hard-interaction events to model the effects from pileup. The detector response to all particles is simulated through \GEANTfour~\cite{GEANT4}, and all final-state objects are reconstructed using the same algorithms used for data.

\section{Corrections and backgrounds \label{section:corrections}}

The MC simulations are corrected to improve the modeling of the data. First, weight factors are applied to all simulated events to match the pileup distribution in data, which consists of roughly 20 interactions per crossing. These weights are based on the measured instantaneous luminosity and the total inelastic cross section that provides a good description of the average number of reconstructed vertices.

The total lepton-selection efficiency is factorized into the product of reconstruction, identification, isolation, and trigger efficiencies, with each component measured in samples of \zll events through a ``tag-and-probe'' method~\cite{Chatrchyan:2012xi,Khachatryan:2015hwa}, in bins of lepton \pt and $\eta$. A charge-dependent efficiency in the muon triggering and reconstruction was observed in previous CMS measurements~\cite{Khachatryan:2016pev}. In the muon channel, all efficiencies are therefore determined separately for positively and negatively charged muons. The same procedures are used for data as for the simulated events, and scale factors are extracted to match the simulated event-selection efficiencies to those in the data.

The lepton momentum is calibrated using \zll events~\cite{Bodek:2012id}. The dominant sources of the mismeasurement of muon momentum originate from the mismodeling of tracker alignment and of the magnetic field. The correction parameters are obtained in bins of muon $\eta$ and $\phi$. First, the average $1/\pt$ values of the reconstructed muon curvature in data and simulation are corrected to the corresponding values calculated for MC generated muons. Then, using MC simulation, the resolution in the reconstructed muon momentum is parametrized as a function of the muon \pt in bins of muon $\abseta$ and the number of tracker hits used in the reconstruction. Next, the correction parameters of the muon momentum scale are fine-tuned by matching the average dimuon mass in each bin of muon charge, $\eta$, and $\phi$ to their reference values. At this point, the ``reference'' distributions, which are based on the generated muons, are smeared by the reconstruction resolution derived in the previous step.
Finally, the scale factors for the muon momentum resolution, in bins of muon $\abseta$, are determined by fitting the ``reference'' dimuon mass distribution to data.

A similar procedure is followed for electrons to reduce the small residual difference between the data and MC simulation. Unlike for muons, the measured electron energy is dominated by the calorimeter, and the corrections are extracted identically for electrons and positrons. The electron energy-scale parameters are fine-tuned by correcting the average dielectron mass in each bin of electron $\eta$ and $\phi$ to the corresponding ``reference'' values. Here, the ``reference'' distributions are based on the generated electrons (post FSR), combined with the FSR photons in a cone, and smeared by the reconstructed energy resolution.

The EW and top quark backgrounds are estimated using MC simulations based on the cross sections calculated at next-to-the-next-to-leading order in QCD~\cite{FEWZ, toppp} and normalized to the integrated luminosity. We use cross sections calculated at NLO for the diboson backgrounds. The multijet background in dimuon events, dominated by muons from heavy-flavor hadron decays, is evaluated using same-sign (SS) dimuon events. A small EW and top quark contamination is evaluated in an MC simulation and subtracted from the SS sample. The distributions are then scaled by roughly a factor of 2, estimated from simulated events, to obtain the multijet contamination in the signal OS dimuon sample.
The multijet background in the dielectron analysis is evaluated using the SS sample in combination with the $\Pe\mu$ events to subtract the contribution from the OS events caused by the misidentification of charge. The distributions used to estimate the background from jets misidentified as leptons (that include the multijet and \PW+jet events) are obtained from the SS $\Pe\mu$ sample. These distributions are used to fit the dielectron mass distribution in the SS events in each $\yll$ bin to extract the normalization of this background.

\begin{figure*}[htbp]
\centering
    \includegraphics[width=\cmsFigWidth]{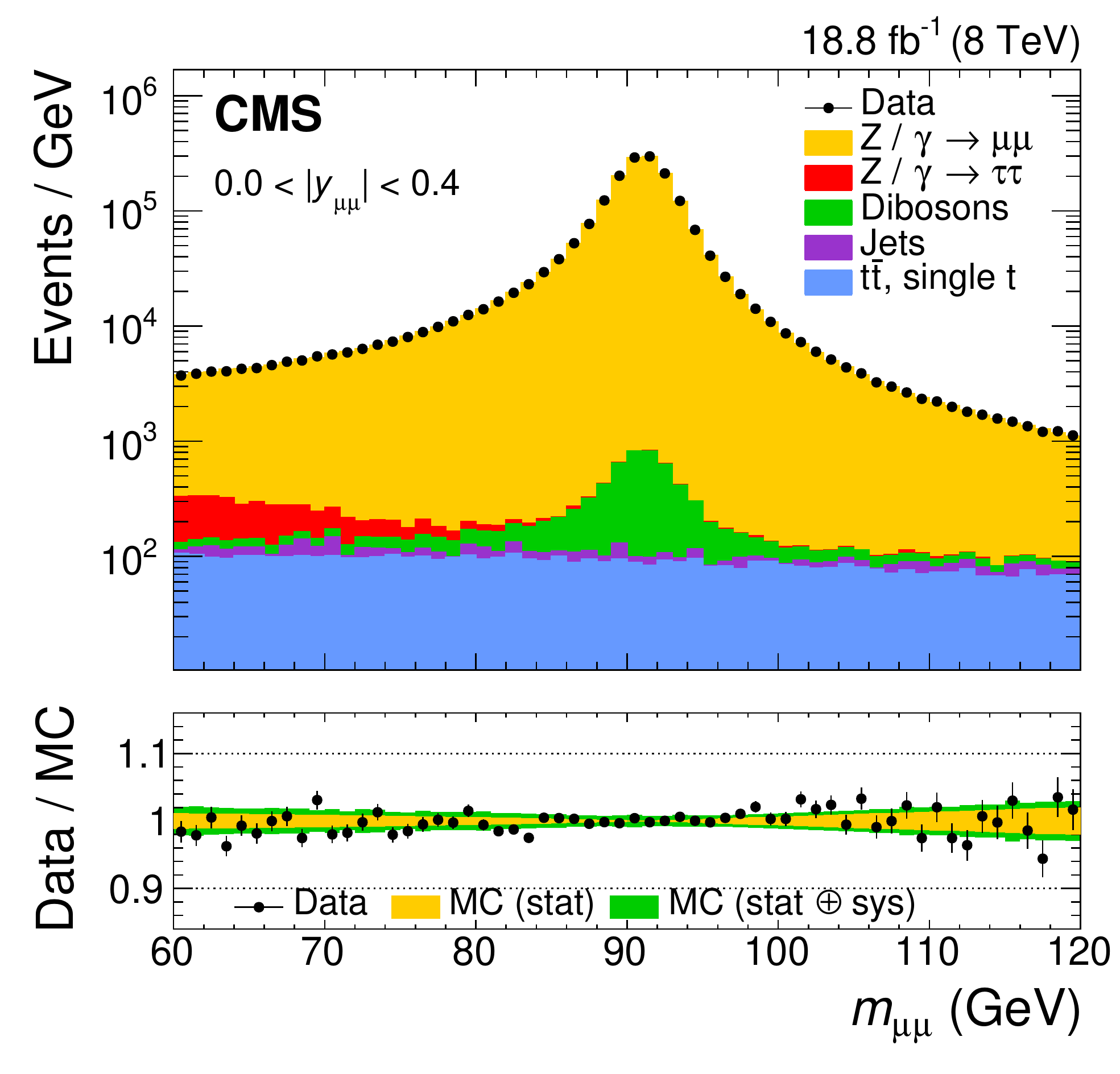}
    \includegraphics[width=\cmsFigWidth]{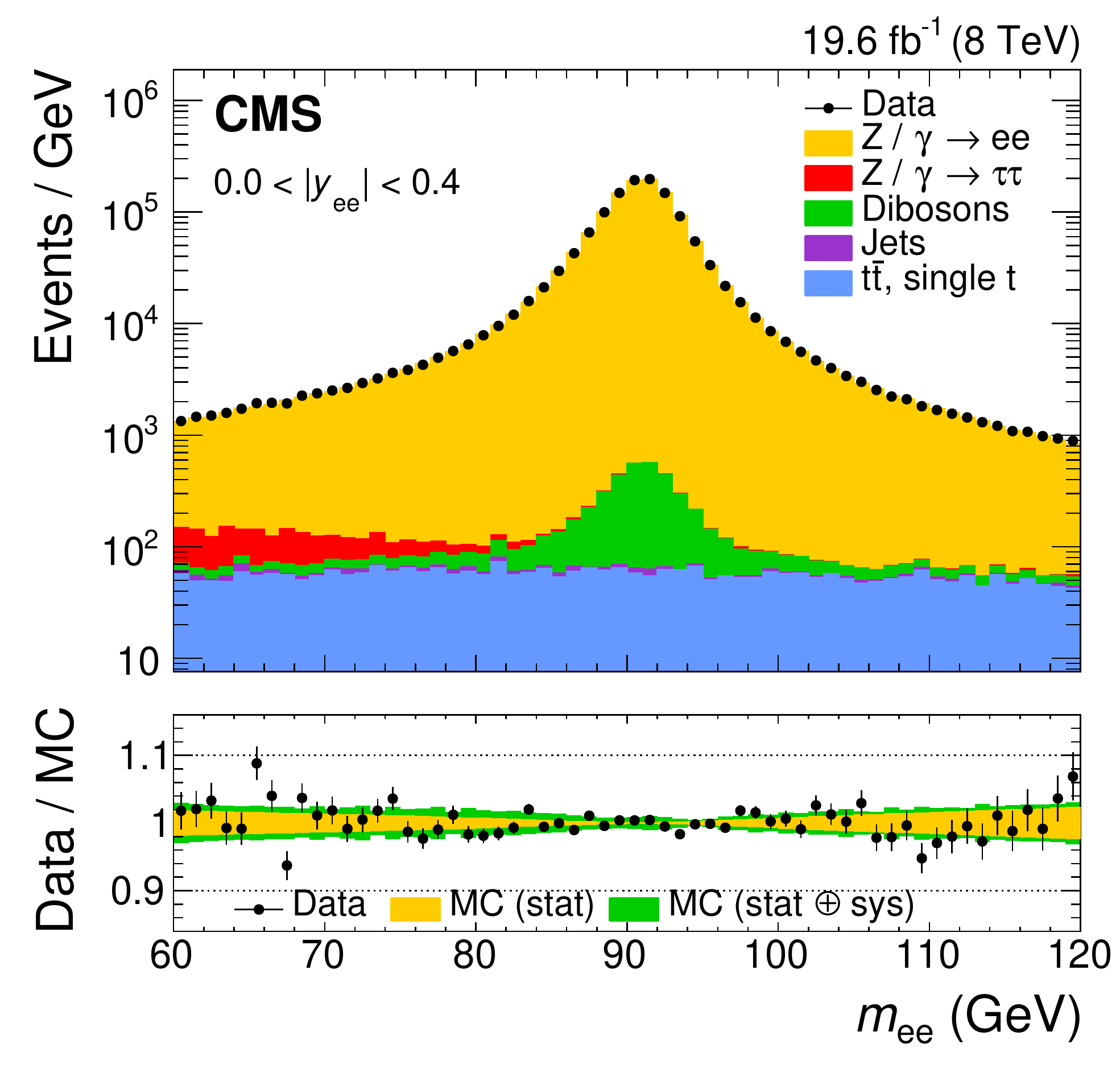}
    \includegraphics[width=\cmsFigWidth]{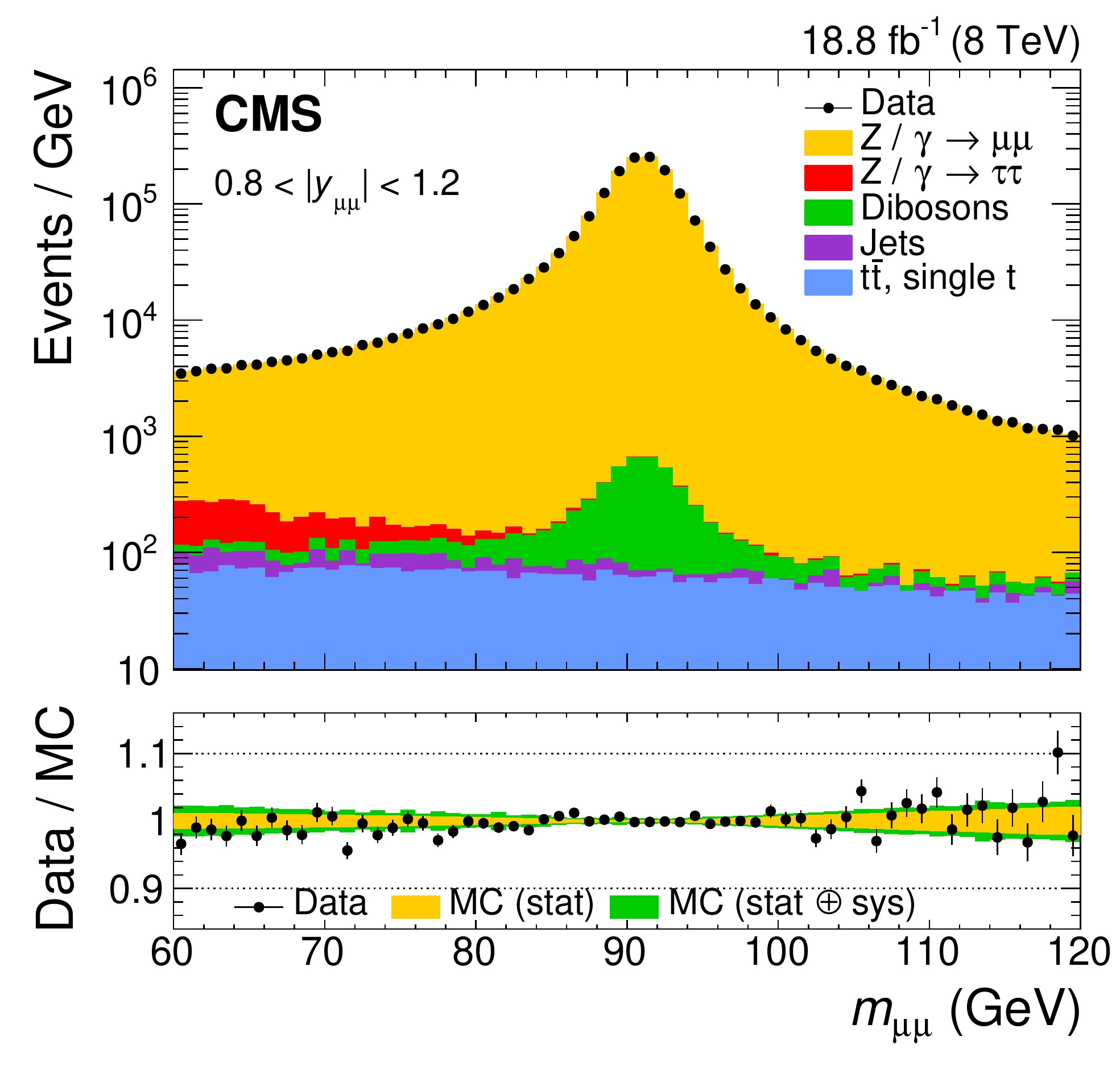}
    \includegraphics[width=\cmsFigWidth]{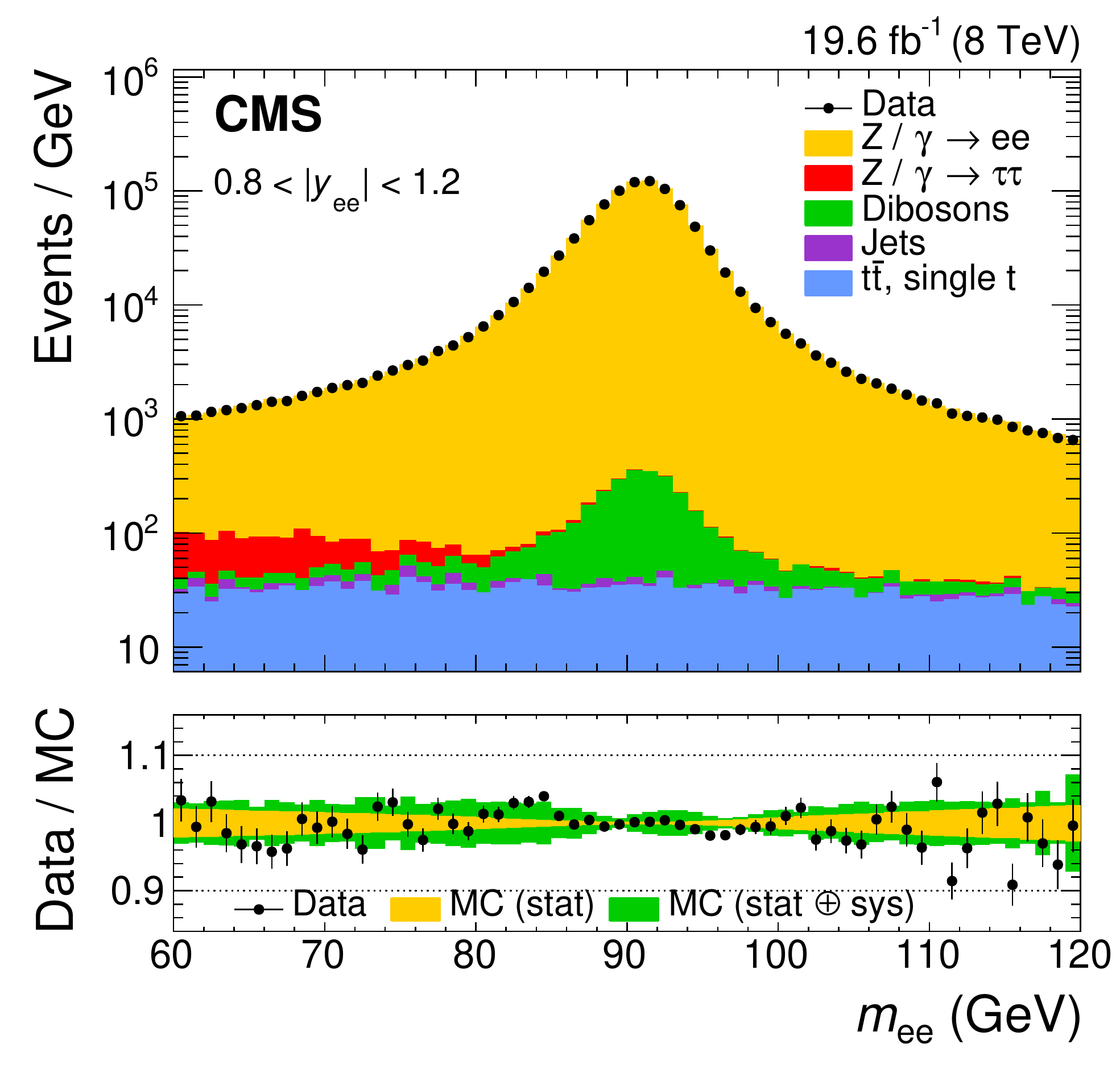}
    \includegraphics[width=\cmsFigWidth]{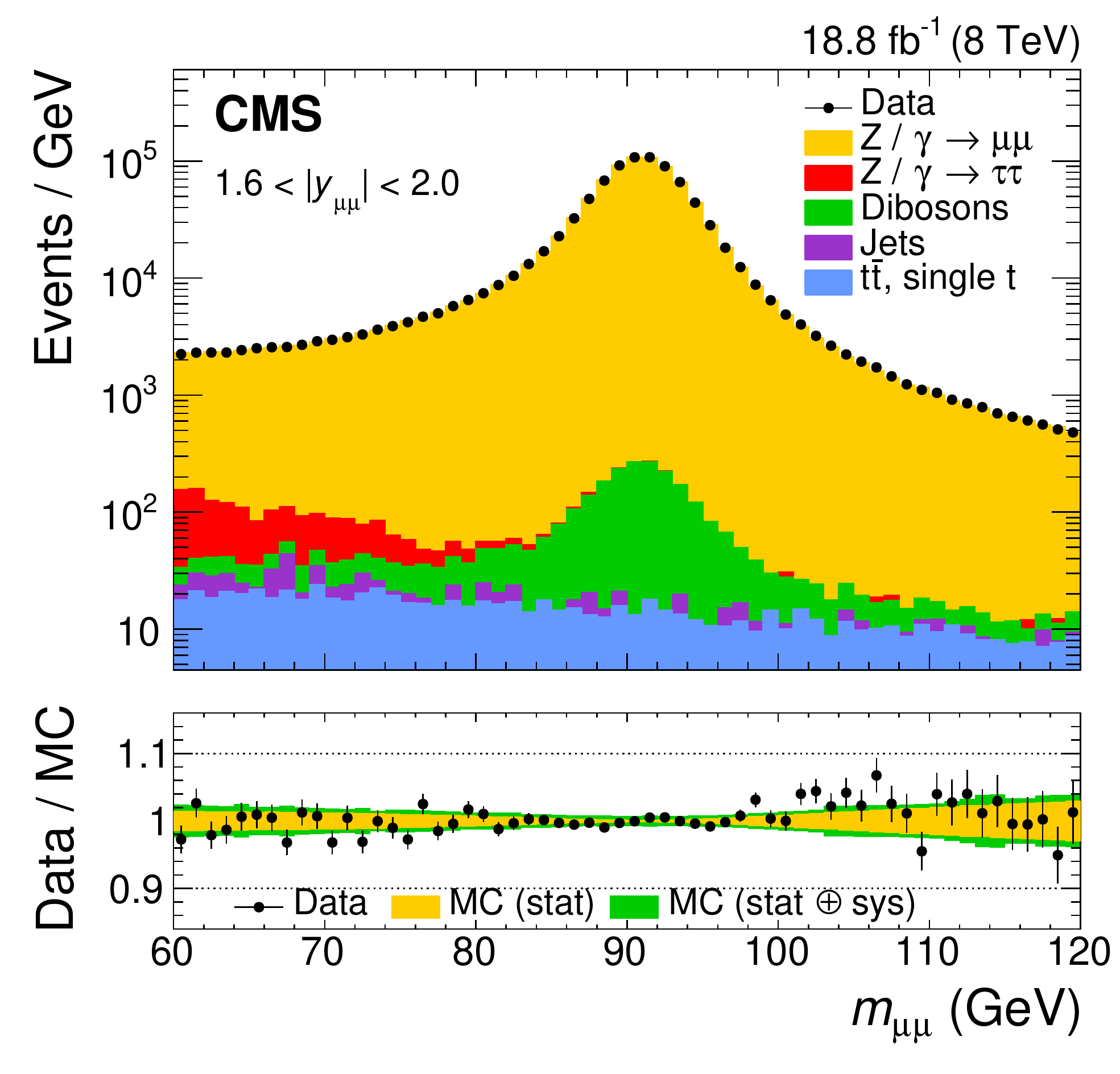}
    \includegraphics[width=\cmsFigWidth]{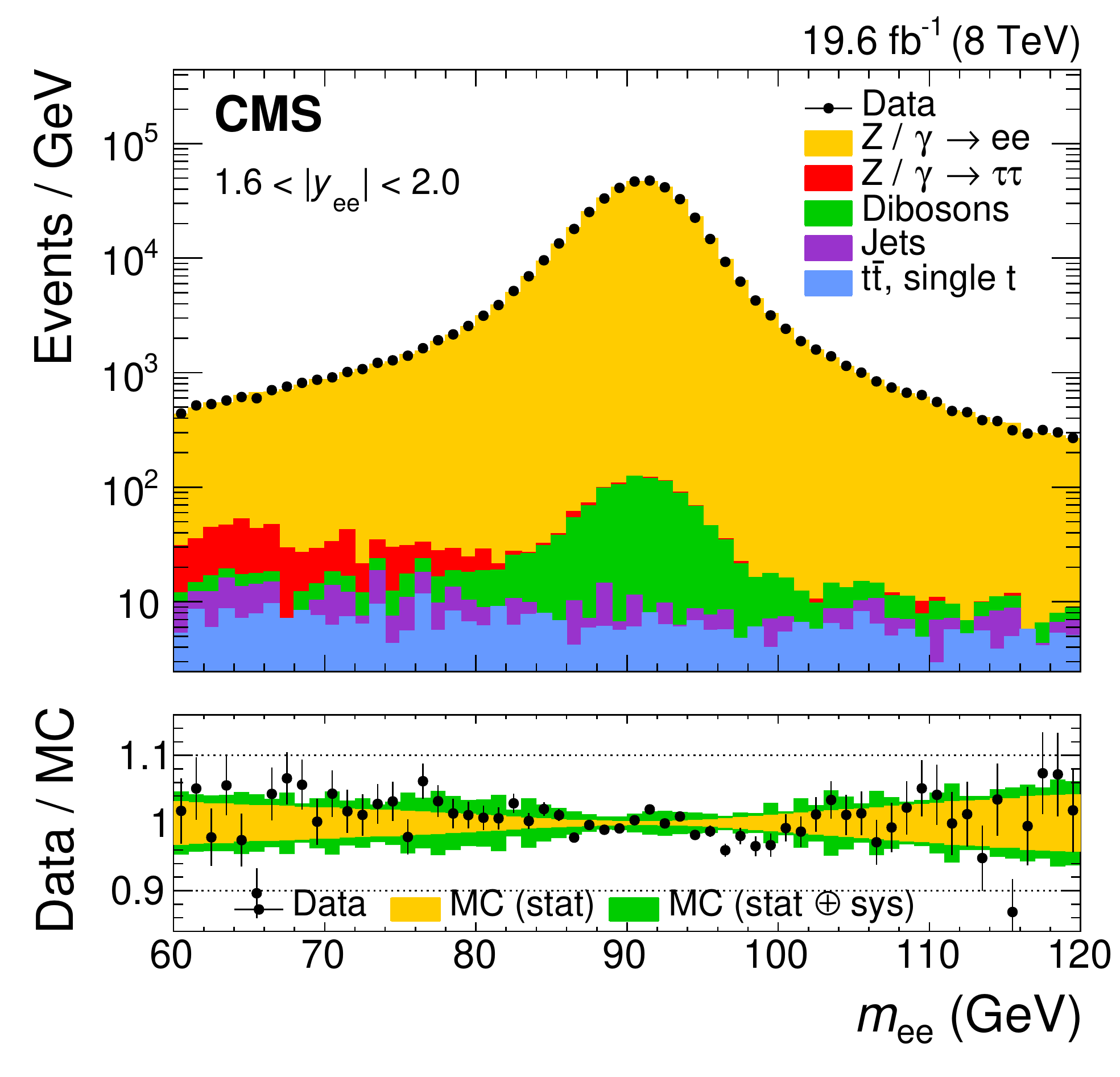}
    \caption{
	Dimuon (left) and dielectron (right) mass distributions in three representative bins in rapidity:
	$\absyll<0.4$ (upper), $0.8<\absyll<1.2$ (middle), and $1.6<\absyll<2.0$ (lower).
	\label{figure:mll}
    }
\end{figure*}

\begin{figure*}[htbp]
\centering
    \includegraphics[width=\cmsFigWidth]{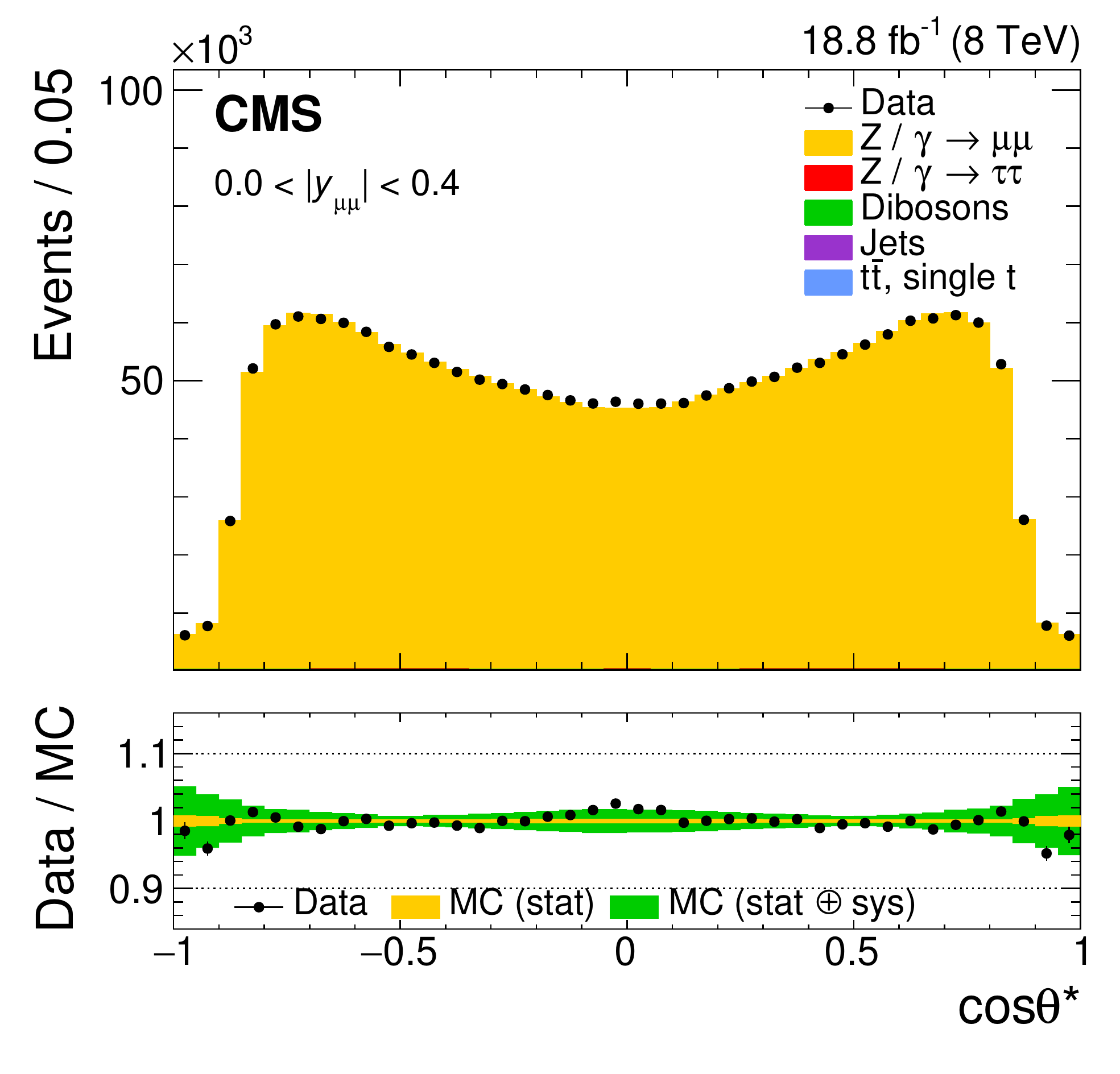}
    \includegraphics[width=\cmsFigWidth]{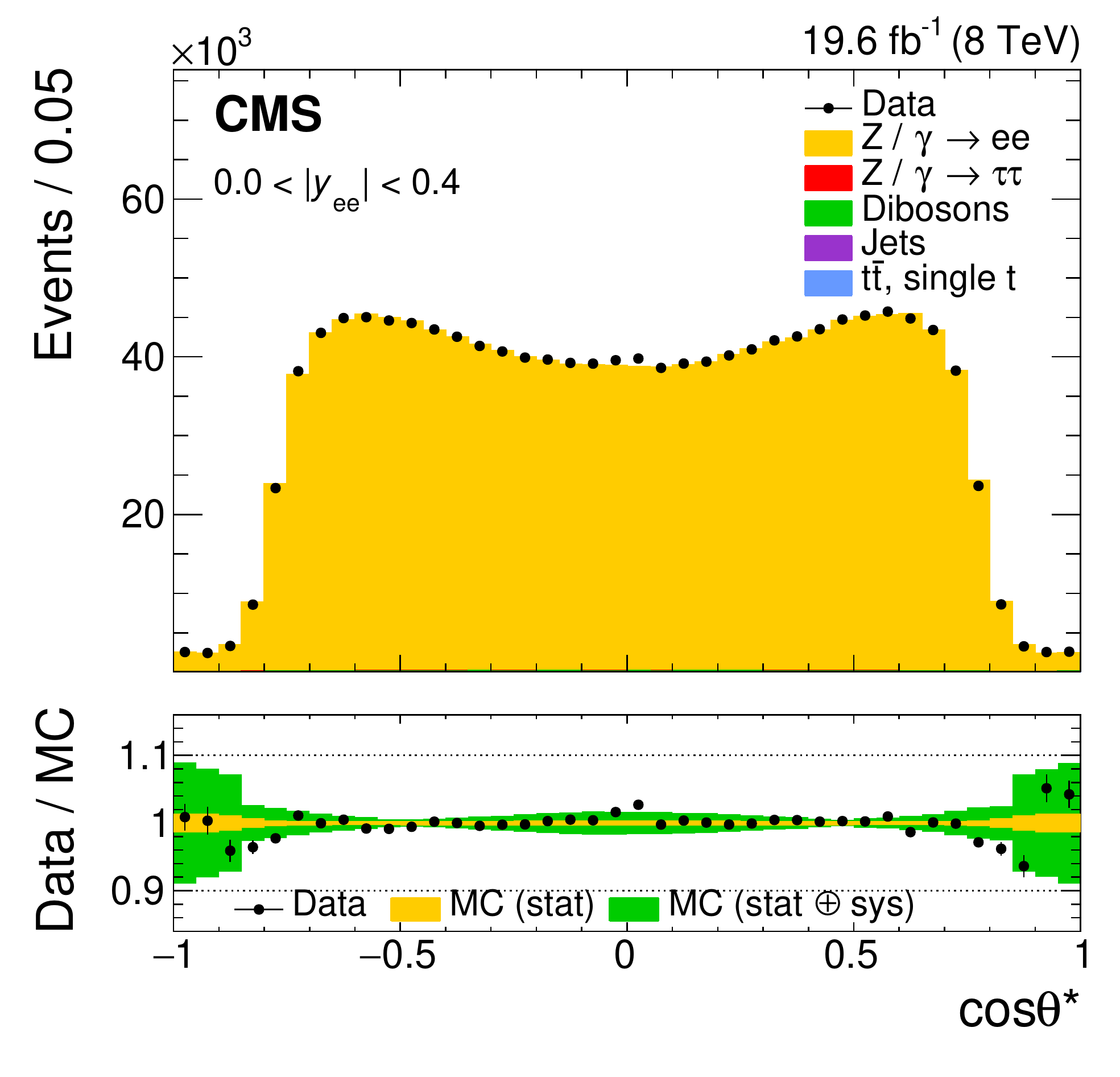}
    \includegraphics[width=\cmsFigWidth]{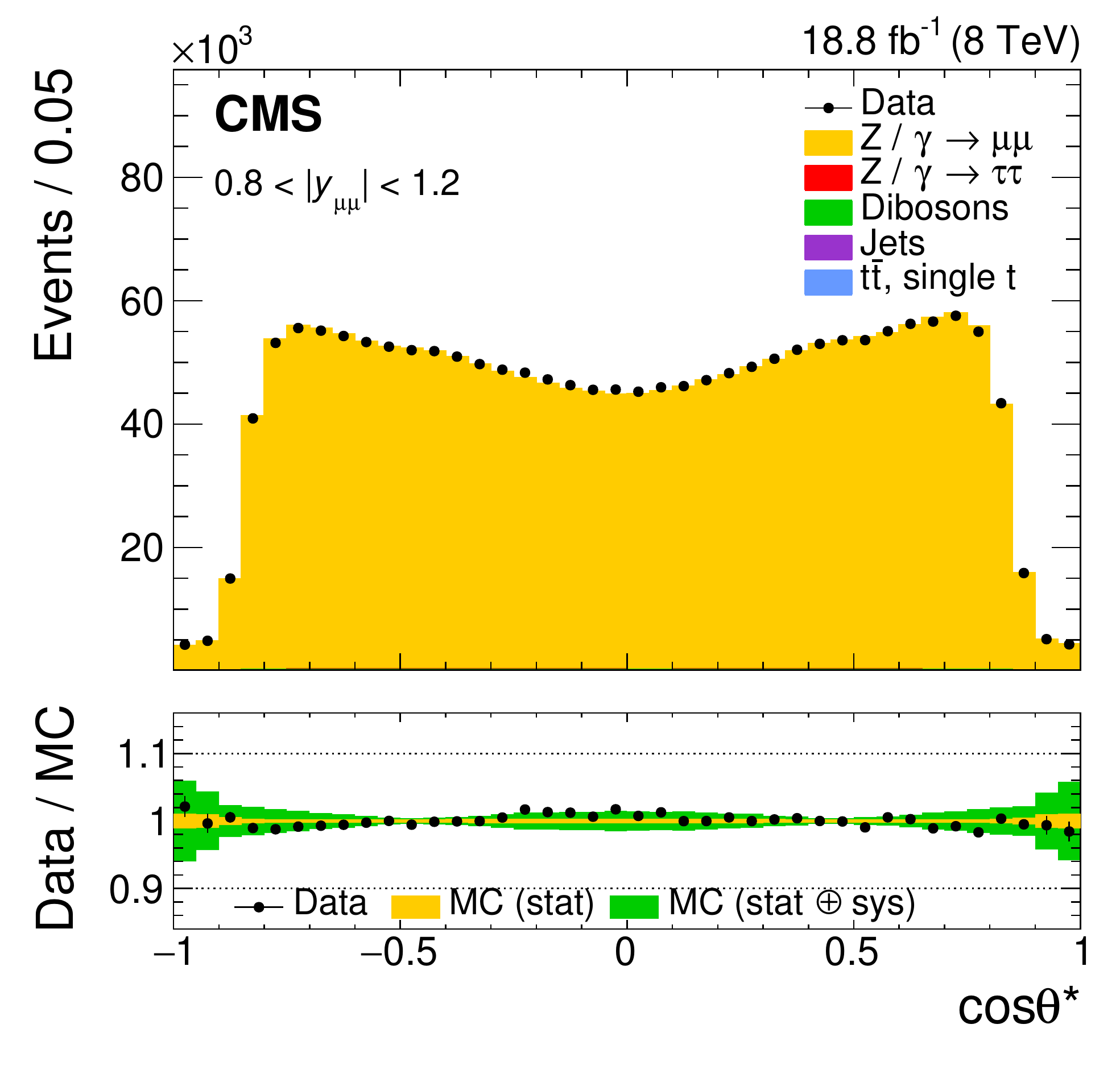}
    \includegraphics[width=\cmsFigWidth]{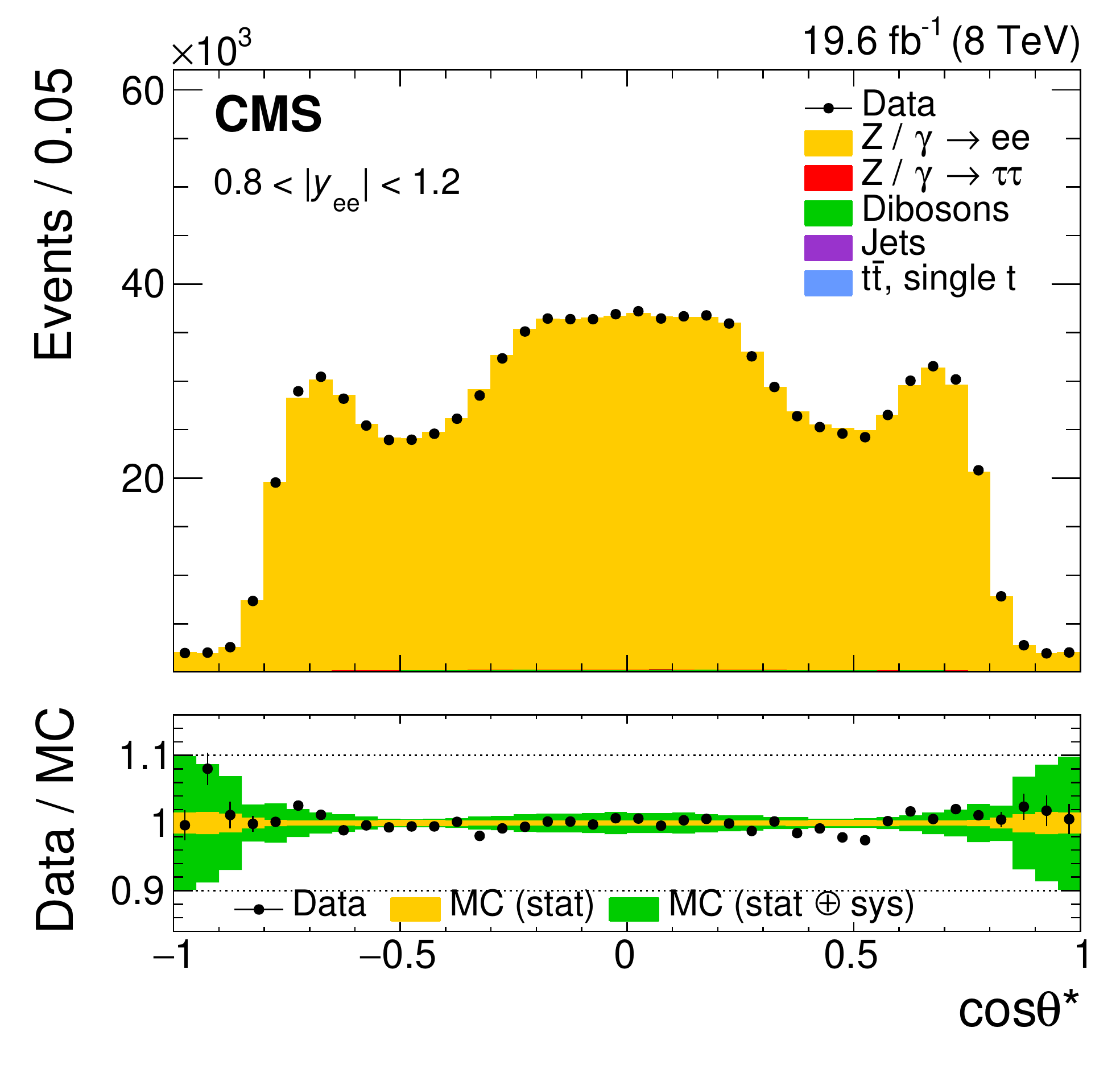}
    \includegraphics[width=\cmsFigWidth]{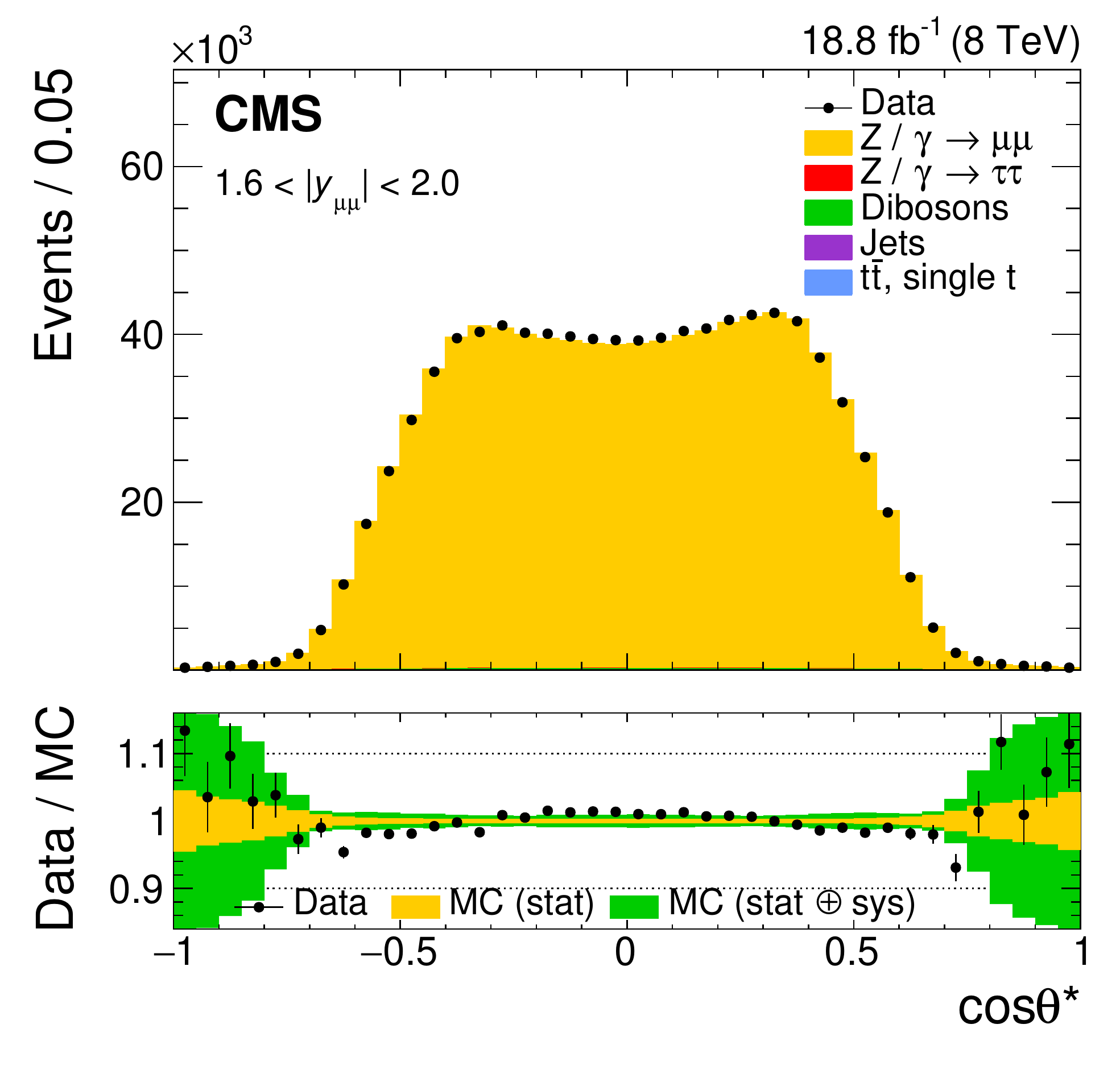}
    \includegraphics[width=\cmsFigWidth]{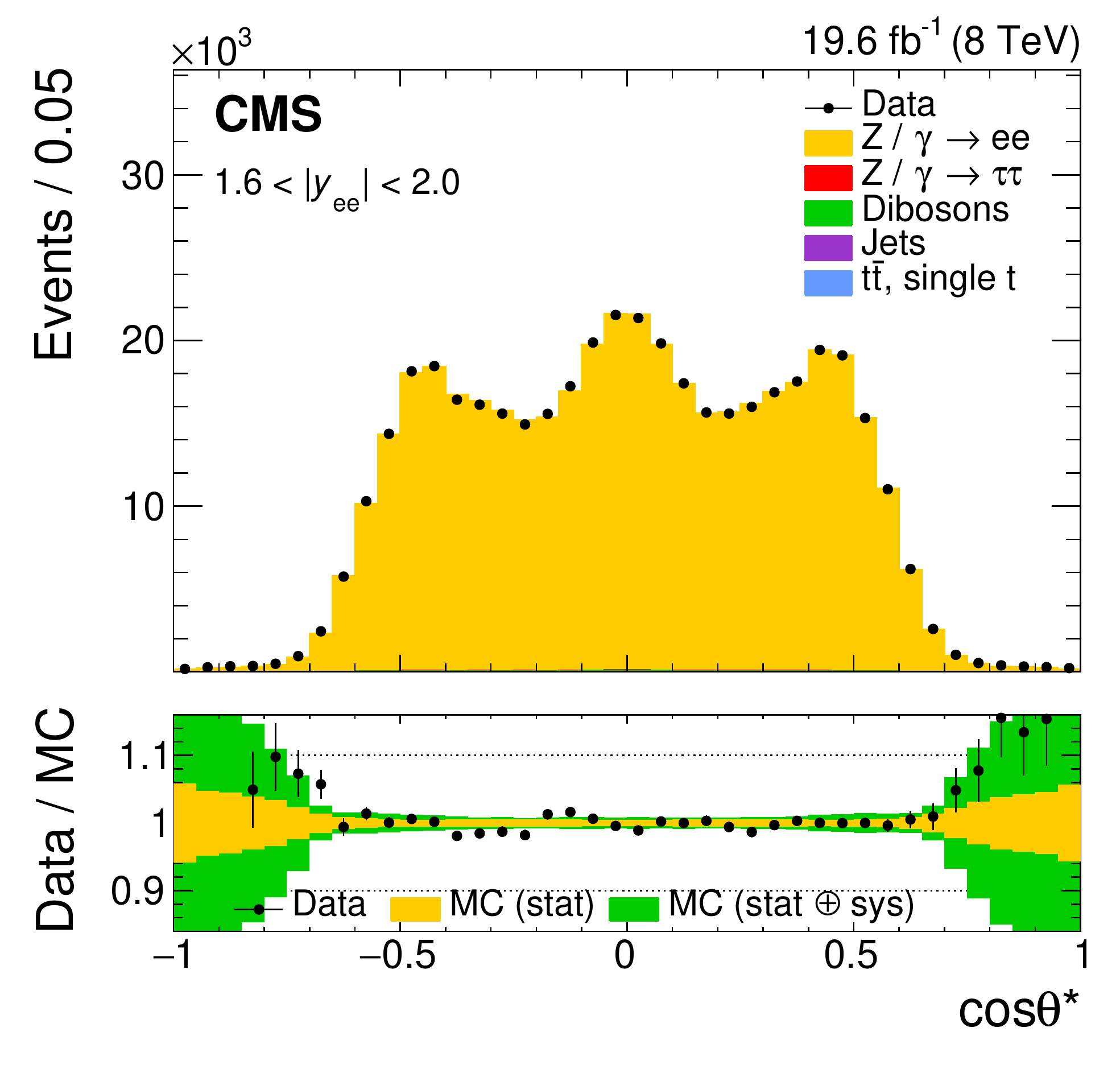}
    \caption{
	The muon (left) and electron (right) \csh distributions in three representative bins in rapidity:
	$\absyll<0.4$ (upper), $0.8<\absyll<1.2$ (middle), and $1.6<\absyll<2.0$ (lower).
	The small contributions from backgounds are included in the predictions.
	\label{figure:mcs}
    }
\end{figure*}

The dilepton mass and \csh distributions in three of the six rapidity bins are shown in Figs.~\ref{figure:mll} and \ref{figure:mcs}, respectively. The figures include lepton momentum and efficiency corrections, background samples normalized as described above, and the signal normalized to the total expected number of events in the data.

\section{Weighted \texorpdfstring{\AFB}{Lg} measurement}

As introduced in Section \ref{introduction}, the LO angular distribution of dilepton events has a $(1+\cos^2\theta^*)$ term that arises from the spin-1 of the exchanged boson and a \csh term that originates from the interference between vector and axial-vector contributions. However, there is also a $(1-3\cos^2\theta^*)$ NLO term that originates from the \pt of the interacting partons~\cite{angular}.
Each $(\mll,\yll)$ bin of the dilepton pair at NLO therefore has an angular distribution in \csh that follows the form~\cite{angular}:
\begin{equation}
    \label{eq:dsigmadcs}
\frac{1}{\sigma}\frac{\rd\sigma}{\rd\csh} = \frac{3}{8}\Big[1+\cos^2\theta^*+\frac{A_0}{2}(1-3\cos^2\theta^*) + A_4\csh\Big].
\end{equation}
The $\AFB$ value in each $(\mll,\yll)$ bin is calculated using the ``angular event weighting'' method, described in Ref.~\cite{Bodek:2010qg}, in which each event with a \csh value (denoted as ``$c$''), is reflected in the denominator ($D$) and numerator ($N$) weights through:
\begin{gather}
    w_\txtD=\frac{1}{2}\frac{c^2}{(1+c^2+h)^3}, \\
    w_\txtN=\frac{1}{2}\frac{\abs{c}}{(1+c^2+h)^2},
\end{gather}
where $h=0.5A_0(1-3c^2)$. Here, as a baseline we use the \ptll-averaged $A_0$ value of about $0.1$ in each measurement $(\mll,\yll)$ bin, as predicted by the signal MC simulation. Using the weighted sums $N$ and $D$ for forward ($\csh>0$) and backward ($\csh<0$) events, we obtain
\begin{gather}
    D_\txtF=\sum_{c>0}w_\txtD, \quad
    D_\txtB=\sum_{c<0}w_\txtD, \\
    N_\txtF=\sum_{c>0}w_\txtN, \quad
    N_\txtB=\sum_{c<0}w_\txtN,
\end{gather}
from which the weighted \AFB of Eq.~(\ref{eq:afb}) can be written as:
\begin{equation}
    \AFB=\frac{3}{8}\frac{N_\txtF-N_\txtB}{D_\txtF+D_\txtB} \label{eq:weightedafb}.
\end{equation}

The statistical uncertainty in this weighted \AFB value takes into account correlations among the numerator and denominator sums. For data, the background contribution in the event-weighted sums are subtracted before calculating \AFB. In the full phase space, the values of the weighted and the nominal \AFB, calculated as an asymmetry between the total event counts in the forward and backward hemispheres, are the same. Since the acceptances of the forward and backward events are equal for same values of $\abs{\csh}$, the fiducial values of the event-weighted \AFB are also the same as in the full phase space, while the nominal \AFB values are smaller because of the limited acceptance at large \csh. This feature makes an event-weighted \AFB less sensitive than the nominal \AFB to the specific modeling of the acceptance. In addition, because the event-weighted \AFB exploits the full distribution in \csh, as opposed to only its sign in the nominal \AFB, it therefore provides a smaller statistical uncertainty.

\section{Extraction of \texorpdfstring{\sineff}{Lg} \label{section:sineff}}

{\tolerance=1600
We extract \sineff  by fitting the \AFB$(\mll,\yll)$ distribution in data with the theoretical predictions.
The default signal distributions are based on the \POWHEG~v2 event generator using the NNPDF3.0 PDFs~\cite{NNPDF30}. The \POWHEG generator is interfaced with \PYTHIA~8~\cite{PYTHIA8} and the CUETP8M1~\cite{Khachatryan:2015pea} underlying event tune to provide parton showering and hadronization, including electromagnetic FSR. The dependence on \sineff, on the renormalization and factorization scales, and on the PDFs is modeled through the \POWHEG MC generator that provides matrix-element-based, event-by-event weights for each change in these parameters. The distributions are modified to different values of \sineff by weighting each event in the full simulation by the ratio of \csh distributions obtained with the modified and default configurations in each $(\mll,\yll)$ bin. The uncertainties in the simulation of the detector have a small effect because \AFB is extracted through the angular event-weighting technique that is insensitive to efficiency and acceptance.
\par}

\begin{figure*}[htbp]
\centering
    \includegraphics[width=\cmsFigWidthBig]{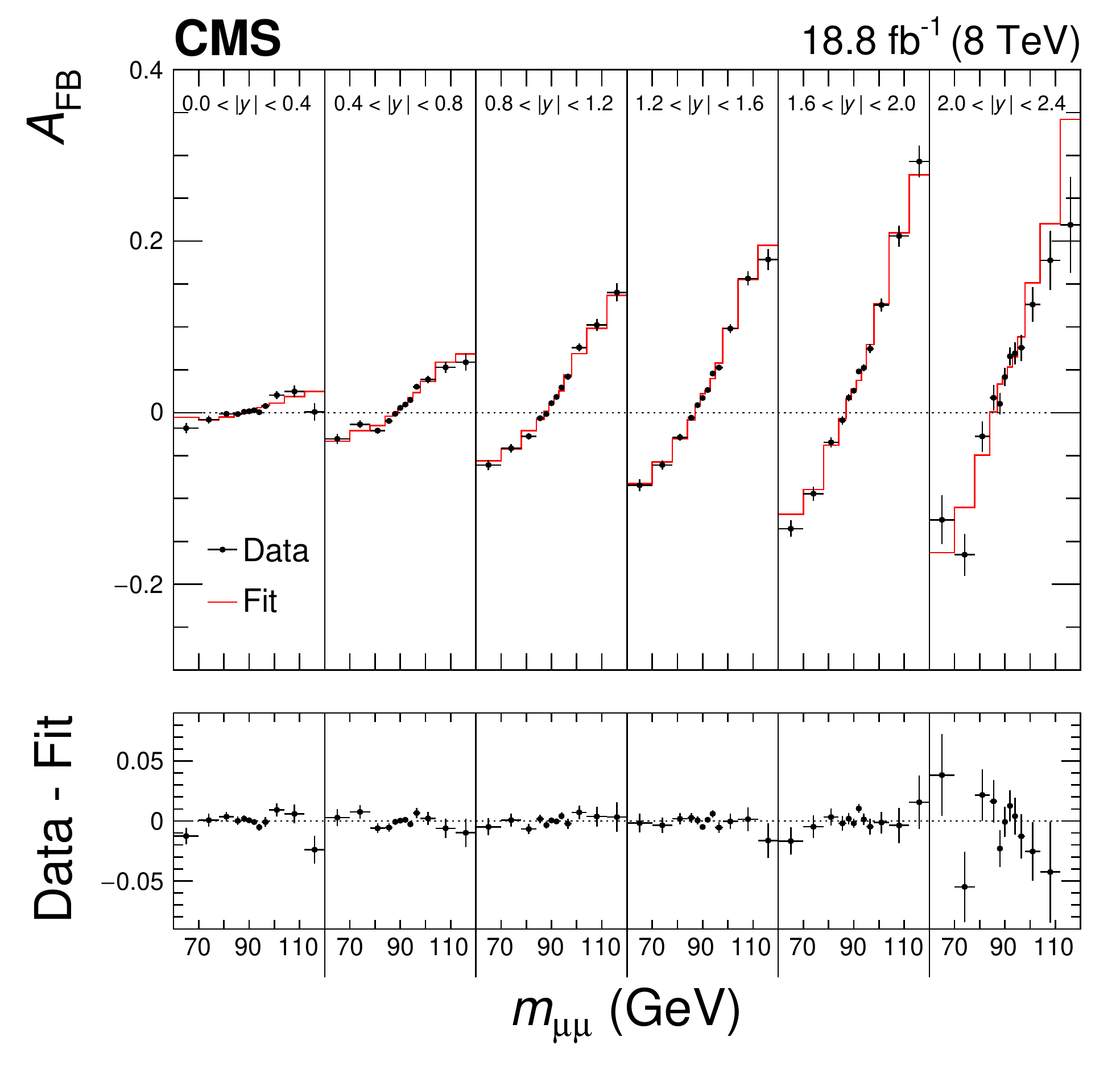}
    \includegraphics[width=\cmsFigWidthBig]{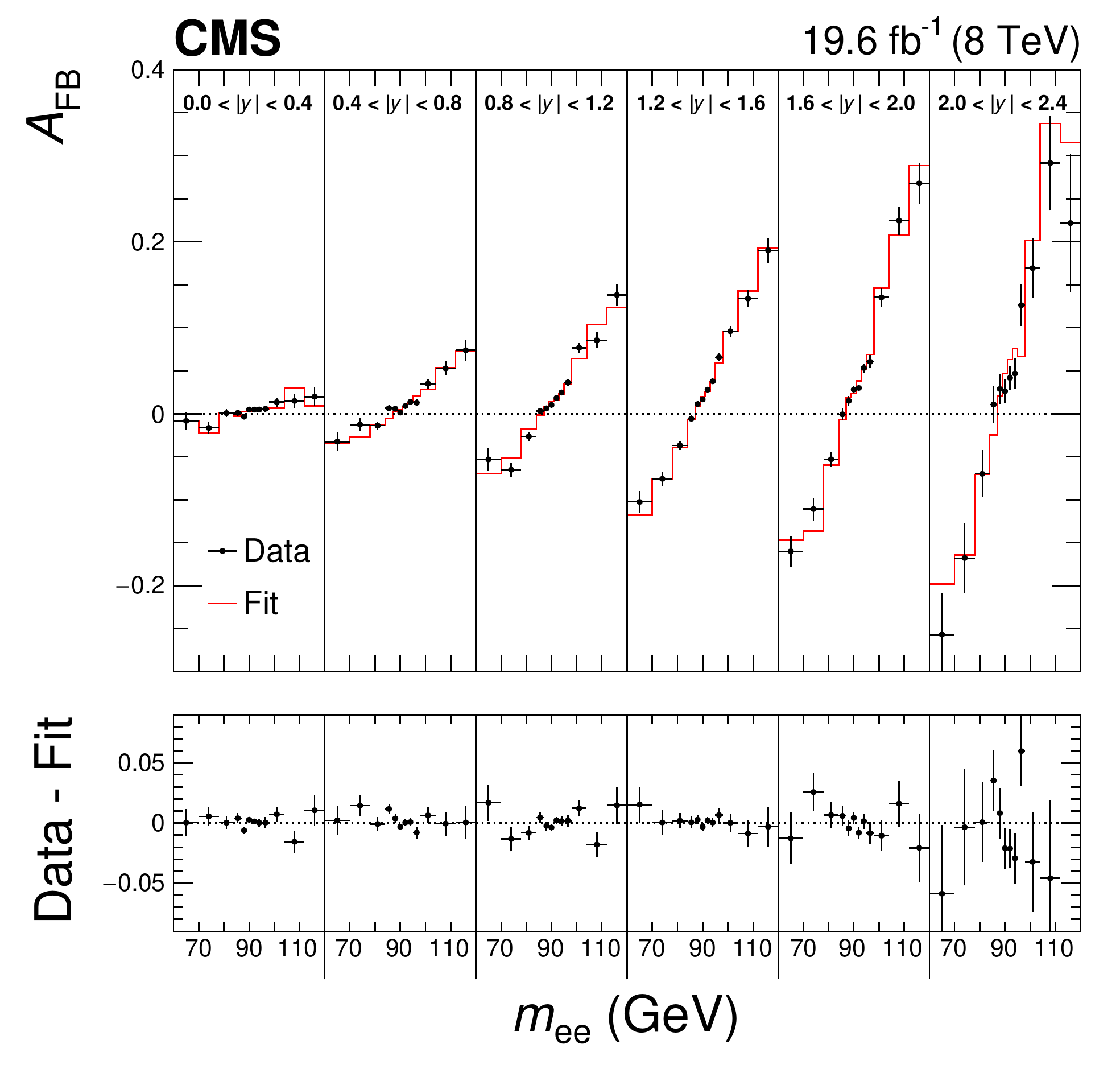}
    \caption{
	Comparison between data and best-fit \AFB distributions in the dimuon (upper) and dielectron (lower) channels.
	The best-fit \AFB value in each bin is obtained via linear interpolation between two neighboring templates.
	Here, the templates are based on the central prediction of the NLO NNPDF3.0 PDFs.
	The error bars represent the statistical uncertainties in the data. \label{figure:fit}
    }
\end{figure*}

{\tolerance=1200
Table~\ref{table:staterrors} summarizes the statistical uncertainty in the extracted \sineff in the muon and electron channels and in their combination. Comparisons between the data and best-fit distributions are shown in Fig.~\ref{figure:fit}.
The statistical uncertainties are evaluated through the bootstrapping technique~\cite{efron1979}, and take account of correlations among the measured \AFB, lepton selection efficiencies, and calibration coefficients introduced through the repeated use of the same dilepton events.
We generate 400 pseudo-experiments that provide an accurate estimate of the statistical uncertainties and correlations. In each pseudo-experiment, every event in the data is replicated $n$ times, where $n$ is a random number sampled from a Poisson distribution with a mean of unity. All steps of the analysis, including extraction of muon selection efficiencies, calibration coefficients, and a measurement of \AFB, are performed for each pseudo-experiment. The statistical uncertainties in electron-selection efficiencies and calibration coefficients, which have no charge dependence, are small and are evaluated separately.
\par}

\begin{table}[htbp]
\centering
\topcaption{
    Summary of statistical uncertainties in \sineff.
    The statistical uncertainties in the lepton-selection efficiency and in the calibration coefficients in data are included in the estimates.
    \label{table:staterrors}
}
\begin{tabular}{  l  c   }
Channel    &  Statistical uncertainty \\ \hline
Muons		       &  0.00044  \\
Electrons	       &  0.00060  \\ [\cmsTabSkip]
Combined	       &  0.00036  \\
\end{tabular}
\end{table}

\section{Experimental systematic uncertainties}

The experimental sources of systematic uncertainty reflect the statistical uncertainties in the simulated events, corrections to lepton-selection efficiency, and to the lepton-momentum scale and resolution, background subtraction, and modeling of pileup. For electrons, the selection efficiencies, which have no dependence on charge, cancel to first order, since we are using the angular event-weighting technique.

\subsection{Statistical uncertainties in  MC simulated events}

{\tolerance=1200
To reduce the statistical uncertainties associated with the limited number of events in the signal MC samples, which include simulation of detector response and lepton reconstruction, the generated \csh distributions in each $(\mll,\yll)$ bin within the acceptance of the detector is reweighted to much larger MC samples, generated without simulating detector response or lepton reconstruction. This makes the fluctuations in the generated \csh distributions negligible, and therefore the statistical uncertainties in the reconstructed \AFB values become dominated by fluctuations in the simulated detector response and lepton reconstruction. These uncertainties are evaluated using the bootstrapping~\cite{efron1979} method in both dimuon and dielectron channels, described in Section~\ref{section:sineff}, by reweighting the generated \csh distributions in each of the bootstrap samples. The total statistical uncertainties in the simulated events also include contributions from uncertainties in the measured lepton-selection efficiencies and calibration coefficients.
\par}

\subsection{Lepton selection efficiencies}

Several sources of uncertainty are considered in measuring of efficiencies. The statistical uncertainties in the lepton-selection efficiencies, evaluated through studies of pseudo-experiments, are included in the combined statistical uncertainty of the measured \sineff.

{\tolerance=800
Combined scale factors for muon reconstruction, identification, and isolation efficiencies are changed by 0.5\%, and trigger-selection efficiency scale factors by 0.2\%, coherently for all bins for both positive and negative lepton charges. These take into account uncertainties associated with the tag-and-probe method, and are evaluated by changing  signal and background models for dimuon mass distributions, levels of backgrounds, the dimuon mass range, and binning used in the fits. These uncertainties are considered fully correlated between the two charges, and therefore have a negligible impact on the measurement of \sineff.
In addition, we assign the difference between the offline efficiencies obtained by fitting the dimuon mass distributions to extract the signal yields, and those found using simple counting method, as additional systematic uncertainties. The total systematic uncertainty in \sineff originating from the muon selection efficiency is $\pm$0.00005.

In a similar way as for muons, the scale factors for electron reconstruction, identification, and trigger-selection efficiencies are changed coherently within their uncertainties in all $(\pt,\eta)$ bins, and the corresponding changes in the resulting \sineff are assigned as systematic uncertainties. The total uncertainty in \sineff originating from all electron efficiency-related systematic sources is $\pm$0.00004.
\par}

\subsection{Lepton momentum calibration}

The statistical uncertainties in the parameters used to calibrate lepton momentum, described in Section~\ref{section:corrections}, are included in the combined statistical uncertainty. The theoretical uncertainties, discussed in Section~\ref{section:theory}, are also propagated to the reference distributions used to extract the coefficients in the lepton momentum calibration.

{\tolerance=1200
When evaluating the average dimuon masses to extract the $(\eta,\phi)$ dependent corrections, the dimuon mass window is restricted to $86<m_{\mu\mu}<96\GeV$. This range of $\pm5\GeV$ centered at 91\GeV is changed from $\pm2.5$ to $\pm10\GeV$ in steps of 0.5\GeV, and the full calibration sequence is repeated each time. Similarly, a dimuon mass window of $\pm10$ (\ie, 81--101)\GeV, used in the dimuon fits to obtain the resolution-correction factors, is changed from $\pm5$ to $\pm25\GeV$ in steps of 1\GeV. For each of these modifications, the maximum deviation in the extracted \sineff relative to the nominal configuration is taken as a systematic uncertainty. The total experimental systematic uncertainty in \sineff originating from the muon-momentum calibration, evaluated by adding individual uncertainties in quadrature, is $\pm$0.00008. The effects due to PDF uncertainties in the calibration coefficients were found to be negligible. In studies of the impact of the value of \sineff used to generate the reference distributions for muon-momentum calibration over the range of $\Delta\sineff=0.02000$, the extracted result changes at most by $\pm$0.00008 due to the changes made in the muon-calibration parameters. Since the uncertainty in \sineff is much smaller than $\pm0.02000$, we conclude that this effect is negligible.
\par}

Similarly, the windows in the dielectron invariant mass used to extract the electron momentum-correction factors are changed to estimate the corresponding systematic uncertainty. And consider additional independent sources of systematic uncertainty from the modeling of pileup, background estimation, and bias in the dielecton mass-fitting procedure. The size of the EW corrections in the extracted electron energy-calibration coefficients is estimated by modifying reference dielectron mass distributions through the weight factors obtained with \textsc{zgrad}~\cite{Baur}. All these systematic uncertainties are found to be rather small. The dominant uncertainty originates from the full corrections to the electron energy resolution, which improve the agreement between data and simulated dielectron mass distributions. The total systematic uncertainty in the extracted value of \sineff due to both the electron energy scale and resolution is $\pm$0.00019.

\subsection{Background}
The systematic uncertainties in the estimated background are evaluated as follows. The normalizations of the top quark and $\ztautau$ backgrounds are changed respectively by 10 and 20\%, covering the maximum deviations between the data and simulation observed in the $\Pe\mu$ control region.
The uncertainty in the multijet and \PW+jets background is estimated by changing them by $\pm$100\%.
Changing the diboson background prediction by 100\% provides a negligible change in the result (${<}0.00001$).
Changing all EW and top quark backgrounds by the uncertainty in the integrated luminosity of 2.6\%~\cite{CMS-PAS-LUM-13-001} also produces a negligible change in the result (${<}0.00001$). The total systematic uncertainty in the measured \sineff from the uncertainty in the background estimation is $\pm0.00003$ and $\pm0.00005$ in the dimuon and dielectron channels, respectively.

\subsection{Pileup}

To take into account the uncertainty originating from differences in pileup between data and simulation, we change the total inelastic cross section by $\pm$5\%, and recompute the expected pileup distribution in data. The analysis is repeated and the difference relative to the central value is taken as the systematic uncertainty. These uncertainties are respectively $\pm0.00003$ and $\pm0.00002$ in the dimuon and dielectron channels.

All the above systematic uncertainties are summarized in Table~\ref{table:expsystematics}.

\section{Theoretical systematic uncertainties \label{section:theory}}

We investigate sources of systematic uncertainty in modeling the MC templates.
For each change in the model, we rederive the reference distributions described in Section~\ref{section:corrections} to adjust the lepton momentum calibration coefficients. As a baseline, the signal MC events are weighted to match the \ptll distribution in each $\absyll$ bin in the data. The difference relative to the result obtained without applying the weight factors, which is $0.00003$ in both channels, is assigned as a systematic uncertainty associated with the modeling of \ptll.

\begin{table}[!h]
\centering
\topcaption{
    Summary of experimental systematic uncertainties in \sineff.
}
\label{table:expsystematics}
\begin{tabular}{ l  c  c }
Source & Muons & Electrons \\ \hline
Size of MC event sample      & 0.00015 &	0.00033 \\
Lepton selection efficiency  & 0.00005 &	0.00004 \\
Lepton momentum calibration  & 0.00008 &	0.00019 \\
Background subtraction	     & 0.00003 &	0.00005 \\
Modeling of pileup	     & 0.00003 &	0.00002 \\ [\cmsTabSkip]
Total			     & 0.00018 &	0.00039 \\
\end{tabular}
\end{table}

The renormalization and factorization scales, $\mu_\txtR$ and $\mu_\txtF$, are each changed independently by a factor of 2, up and down, such that their ratio is within $0.5<\mu_\txtR/\mu_\txtF<2.0$. The maximum deviation among these six variants relative to the nominal choice (excluding the two opposite changes) is assigned as a systematic uncertainty associated with the missing higher-order QCD correction terms.

{\tolerance=1200
In addition, we use a multi-scale improved NLO (\textsc{MiNLO}~\cite{MiNLO}) calculation for the \cPZ+1 jet partonic final state (henceforth referred to as ``\cPZ+j''), interfaced with \PYTHIA~8 for parton showering, FSR, and hadronization, to assess the uncertainty from the missing higher-order QCD terms and modeling of the angular coefficients. The \textsc{MiNLO} \cPZ+j process has NLO accuracy for both \cPZ+0 and \cPZ+1 jet events, which provides a better description of the dependence of the angular coefficients on \ptll.
\par}

Systematic uncertainties in modeling electromagnetic FSR are estimated by comparing results obtained with distributions based on \PYTHIA~8 and \PHOTOS~2.15~\cite{Photos1,Photos2,Photos3} for the modeling of FSR. Electroweak effects from the difference between the \cPqu\ and \cPqd\ quarks and leptonic effective mixing angles, are estimated by changing $\sin^2\theta_\txteff^\cPqu$ and $\sin^2\theta_\txteff^\cPqd$ by 0.0001 and 0.0002~\cite{Baur}, respectively, relative to \sineff. The \sineff extracted using the corresponding distributions is shifted by 0.00001.

The underlying event tune parameters~\cite{Khachatryan:2015pea} are changed by their uncertainties, and \sineff is extracted also using the corresponding distributions. The maximum difference from the default tune is taken as the corresponding uncertainty. The systematic uncertainties from these and all the above sources, are summarized in Table~\ref{table:theorysystematics}.

We also separately study the modeling of the  $A_0$ angular coefficient, which is included in the definition of \AFB. As a baseline, the \ptll-averaged $A_0$ value in each measurement $(\mll,\yll)$ bin is used in the definition of the weighted \AFB. Several other options are studied: (i) the LO expression: $A_0=\ptll^2/(\ptll^2+\mll^2)$, (ii) the $\ptll$-dependent $A_0$ in each $(\mll,\yll)$ bin as predicted in the baseline NLO \POWHEG simulation, (iii) the $\ptll$-dependent $A_0$ predicted in the \textsc{MiNLO} \cPZ+j \POWHEG generator, and (iv) $A_0$ set to 0. The same definition is used for data and simulation, and the extracted \sineff is identical within $\pm$0.00002 of the default. In addition, we weight the $\abs{\csh}$ distribution from the \textsc{MiNLO} \cPZ+j MC sample to match the dependence of $A_0$ on \ptll in each $(\mll,\yll)$ bin to the corresponding values of the baseline MC simulation. The change in the resulting \sineff is also negligible.

\begin{table}[!htbp]
\centering
\topcaption{
    \label{table:theorysystematics}
    Summary of the theoretical uncertainties for the dimuon and dielectron channels, as discussed in the text.
}
\label{table:systheorypdf}
\begin{tabular}{ l  c  c  }
Modeling parameter &  Muons &  Electrons \\ \hline
Dilepton \pt reweighting				& 0.00003 &  0.00003 \\
$\mu_\txtR$ and $\mu_\txtF$ scales			& 0.00011 &  0.00013 \\
\POWHEG \textsc{MiNLO} \cPZ+j \vs \PZ at NLO		& 0.00009 &  0.00009 \\
FSR model (\PHOTOS \vs\ \PYTHIA~8)			& 0.00003 &  0.00005 \\
Underlying event					& 0.00003 &  0.00004 \\
    Electroweak $\sineff$ \vs $\sin^2\theta^{\cPqu, \cPqd}_\txteff$  & 0.00001 &  0.00001 \\ [\cmsTabSkip]
Total							& 0.00015 &  0.00017 \\
\end{tabular}
\end{table}

\section{Uncertainties in the PDFs\label{section:pdf}}

The observed \AFB values depend on the size of the dilution effect, as well as on the relative contributions from \cPqu\ and \cPqd\ valence quarks to the total dilepton production cross section. The uncertainties in the PDFs translate into sizable changes in the observed \AFB values. However, changes in PDFs affect the $\AFB(\mll,\yll)$ distribution in a different way than changes in \sineff.

\begin{figure}[!htb]
\centering
    \includegraphics[width=0.48\textwidth]{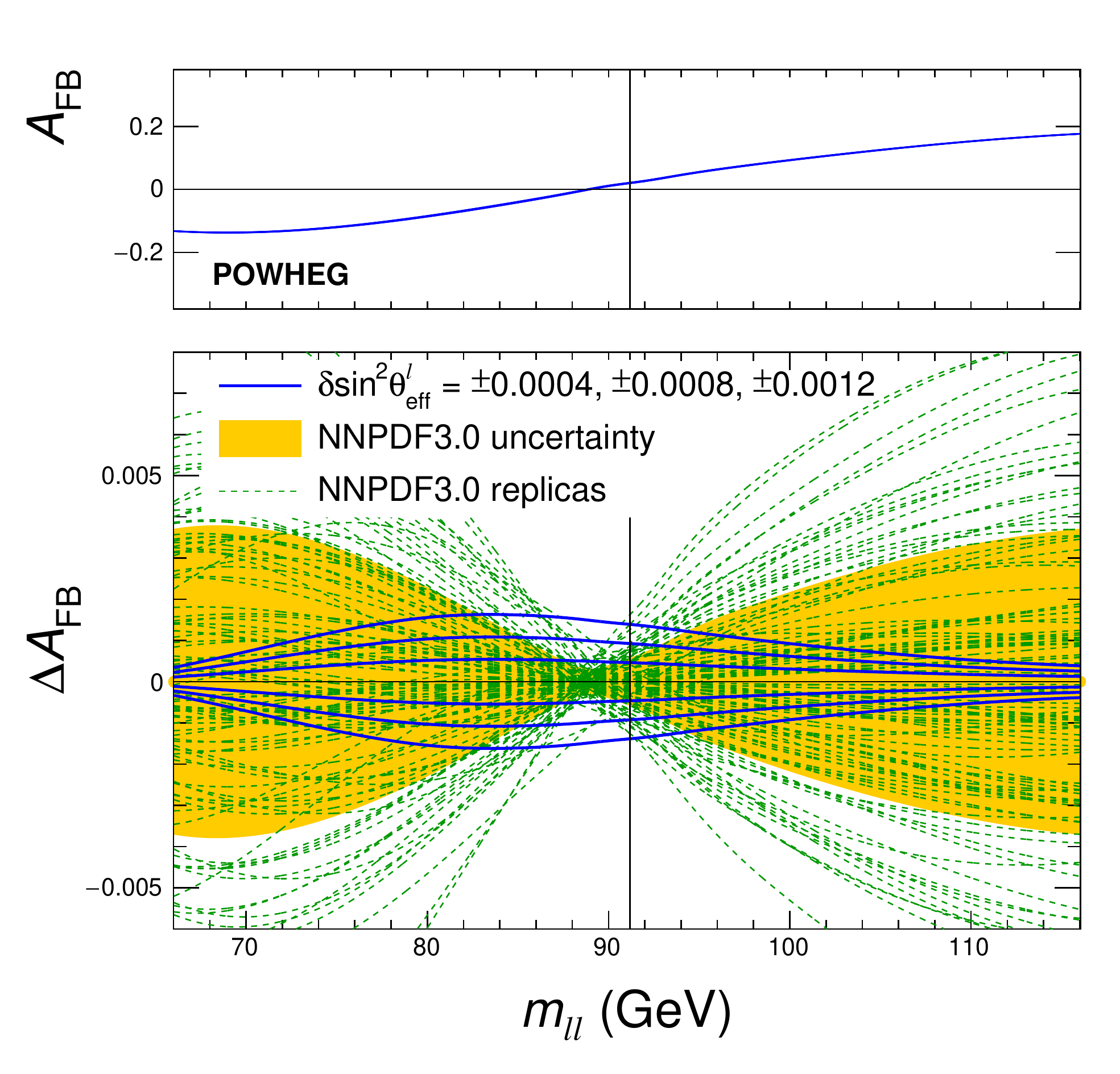}
    \includegraphics[width=0.48\textwidth]{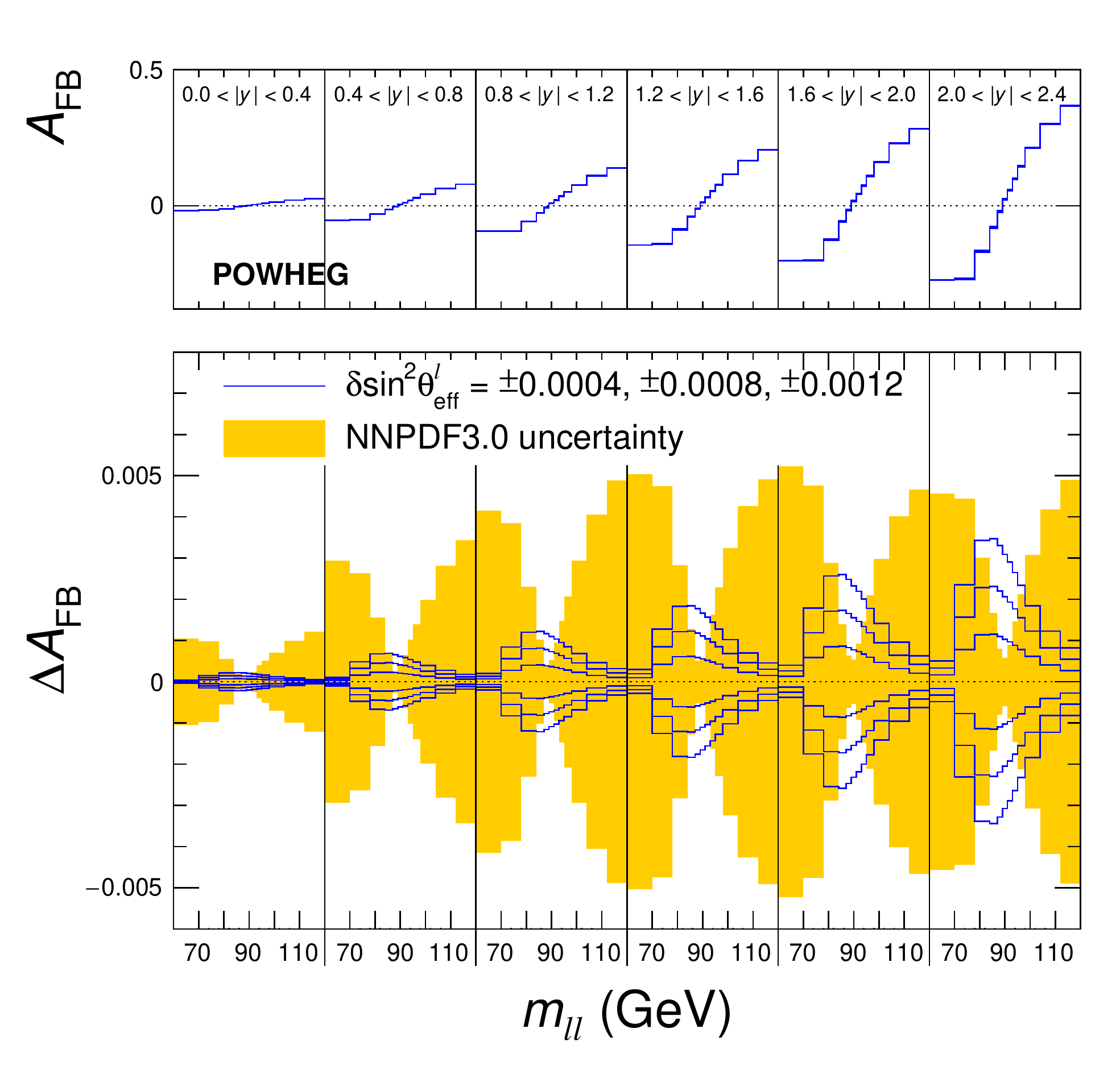}
    \caption{
	Distribution in \AFB as a function of dilepton mass, integrated over rapidity (\cmsLeft),
	and in six rapidity bins (\cmsRight) for $\sineff=0.23120$ in \POWHEG.
	The solid lines in the bottom panel correspond to six changes at $\sineff$ around the central value, corresponding to:
	$\pm0.00040$, $\pm0.00080$, and $\pm0.00120$.
	The dashed lines refer to the \AFB predictions for 100 NNPDF3.0 replicas.
	The shaded bands illustrate the standard deviation in the NNPDF3.0 replicas.
	\label{figure:pdftheory}
    }
\end{figure}

Changes in PDFs produce large changes in \AFB, when the absolute values of \AFB are large, \ie, at large and small dilepton mass values. In contrast, the effect of changes in \sineff are largest near the \PZ boson peak, and are significantly smaller at high and low masses. Because of this behavior, which is illustrated in Fig.~\ref{figure:pdftheory}, we apply a Bayesian $\chi^2$ reweighting method to constrain the PDFs~\cite{Giele:1998gw,Sato:2013ika,Bodek:2016olg}, and thereby reduce their uncertainties in the extracted value of \sineff.

As a baseline, we use the NLO NNPDF3.0 PDFs. In the Bayesian $\chi^2$ reweighting method, PDF replicas that offer good descriptions of the observed \AFB distribution are assigned large weights, and those that poorly describe the \AFB are given small weights. Each weight factor is based on the best-fit $\chi^2_{\txtmin,i}$ value obtained by fitting the \AFB(\mll,\yll) distribution with a given PDF replica $i$:
\begin{equation} \label{eq:bayesweight}
    w_i = \frac{\re^{-\frac{\chi^2_{\txtmin,i}}{2}}}{\frac{1}{N}\sum_{i=1}^N \re^{-\frac{\chi^2_{\txtmin,i}}{2}}},
\end{equation}
where $N$ is the number of replicas in a set of PDFs. The final result is then calculated as a weighted average over the replicas: $\sineff=\sum_{i=1}^{N} w_i s_i/N$, where $s_i$ is the best-fit \sineff value obtained for the $i^\text{th}$ replica.

\begin{figure*}[!htbp]
\centering
    \includegraphics[width=0.48\textwidth]{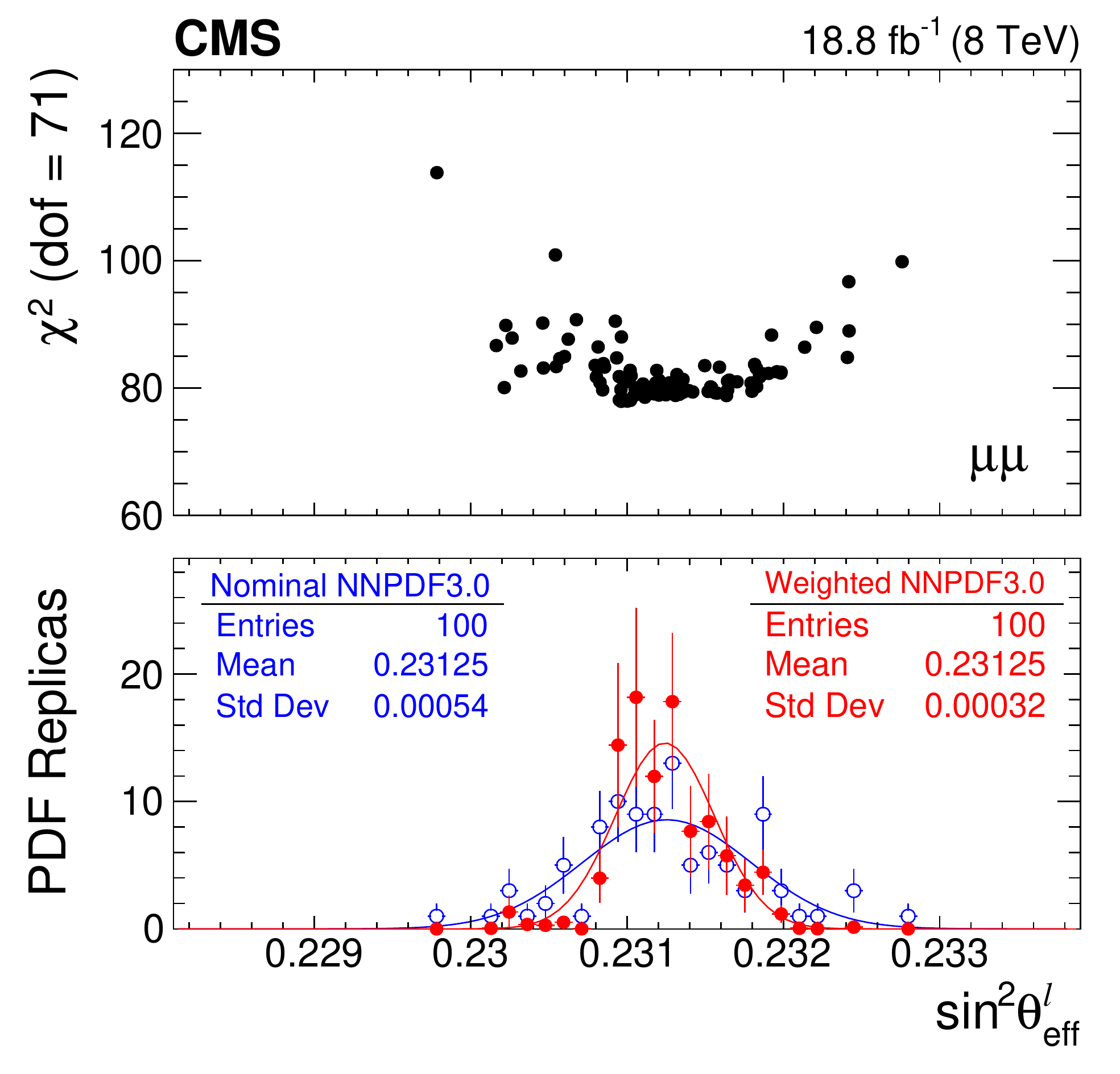}
    \includegraphics[width=0.48\textwidth]{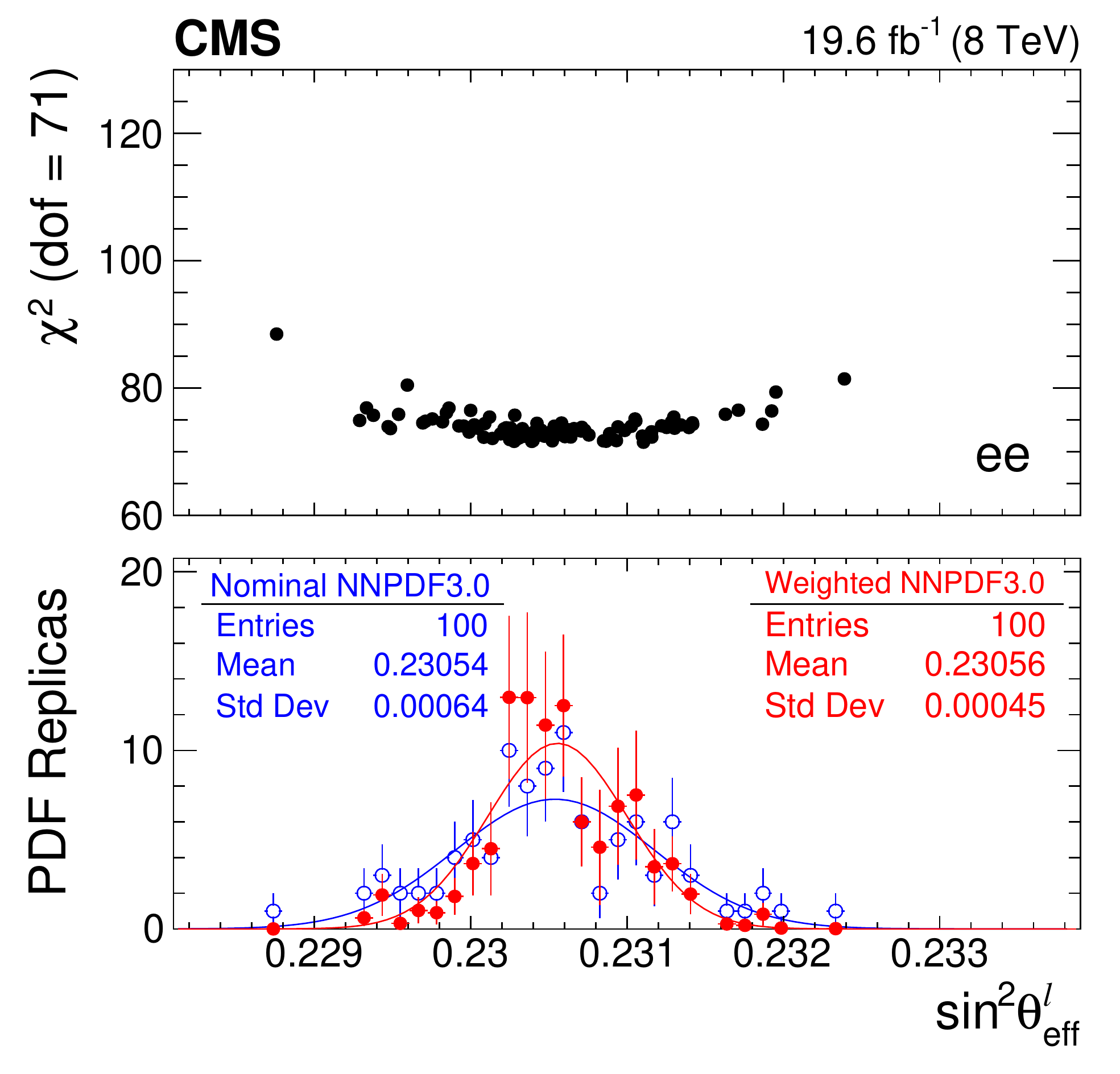}
    \includegraphics[width=0.48\textwidth]{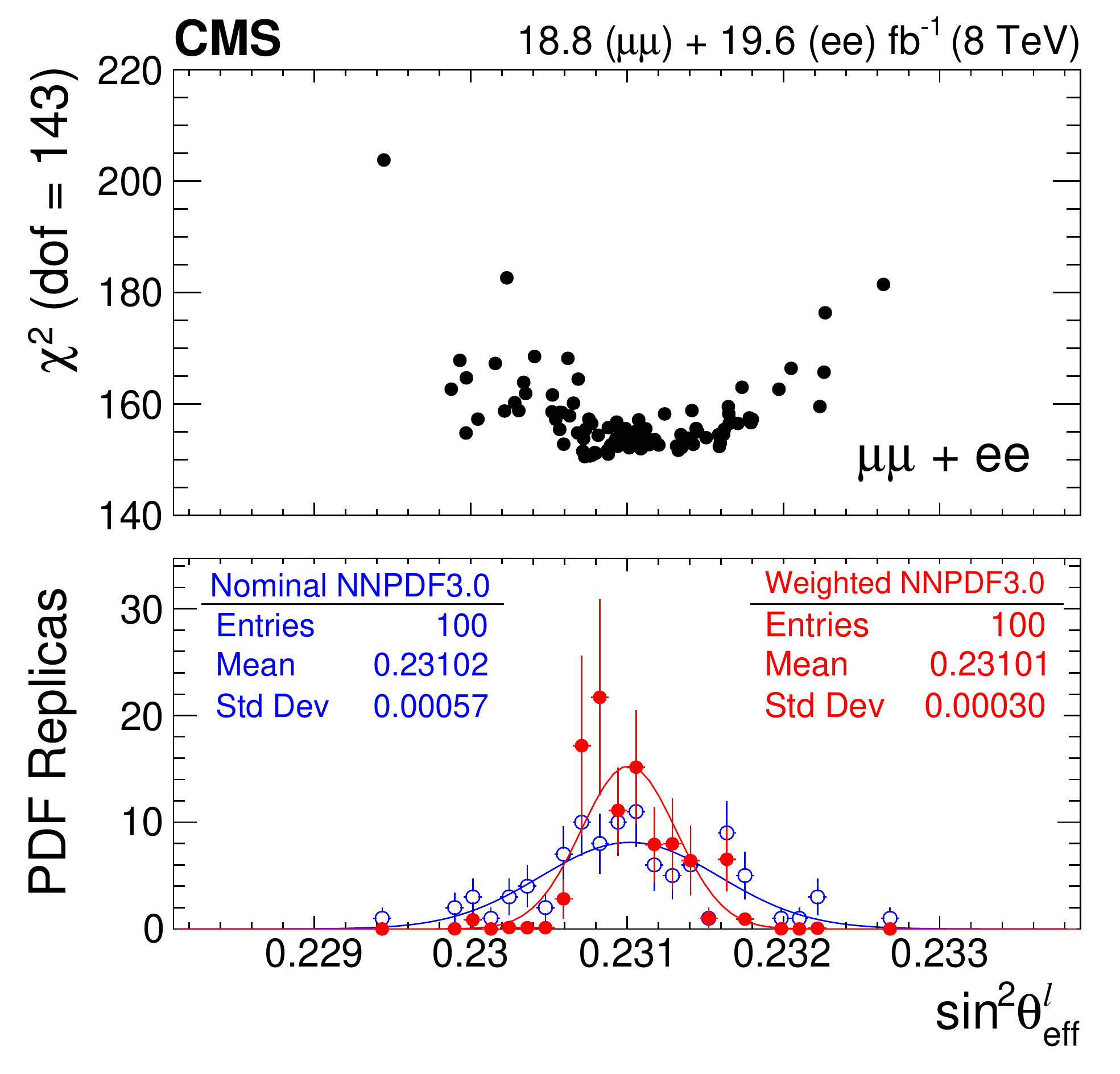}
    \caption{
	The upper panel in each figure shows a scatter plot in $\chi^2_\txtmin$ \vs the best-fit
	\sineff for 100 NNPDF replicas in the muon channel (upper left),
	electron channel (upper right), and their combination (below).
	The corresponding lower panels have the projected distributions in the best-fit \sineff for the nominal (open circles) and weighted (solid circles) replicas.
	\label{figure:comination:scatter}
    }
\end{figure*}

Figure~\ref{figure:comination:scatter} shows a scatter plot of the $\chi^2_{\txtmin}$ \vs the best-fit \sineff value for the 100 NNPDF3.0 replicas for the $\mu\mu$  and $\Pe\Pe$ samples, and for the combined dimuon and dielectron results. All sources of statistical and experimental systematic uncertainties are included in a $72{\times}72$ covariance matrices for data and template \AFB distributions. The $\chi^2(s)$ is defined as:
\begin{equation}
    \chi^2(s)= (\boldsymbol{D}-\boldsymbol{T}(s))^{T}\boldsymbol{V}^{-1}(\boldsymbol{D}-\boldsymbol{T}(s)),
\end{equation}
where $\boldsymbol{D}$ represents the measured \AFB values for data in 72 bins, $\boldsymbol{T}(s)$ denotes the theoretical predictions for \AFB as a function of $s$, or $\sineff$, and $\boldsymbol{V}$ represents the sum of the covariance matrices for the data and templates. As illustrated in these figures, the extreme PDF replicas from either side are disfavored by both the dimuon and dielectron data. For each of the NNPDF3.0 replicas, the muon and electron results are combined using their respective best-fit $\chi^2$ values, \sineff, and their fitted statistical and experimental systematic uncertainties.

\begin{figure}[!htbp]
\centering
    \includegraphics[width=0.48\textwidth]{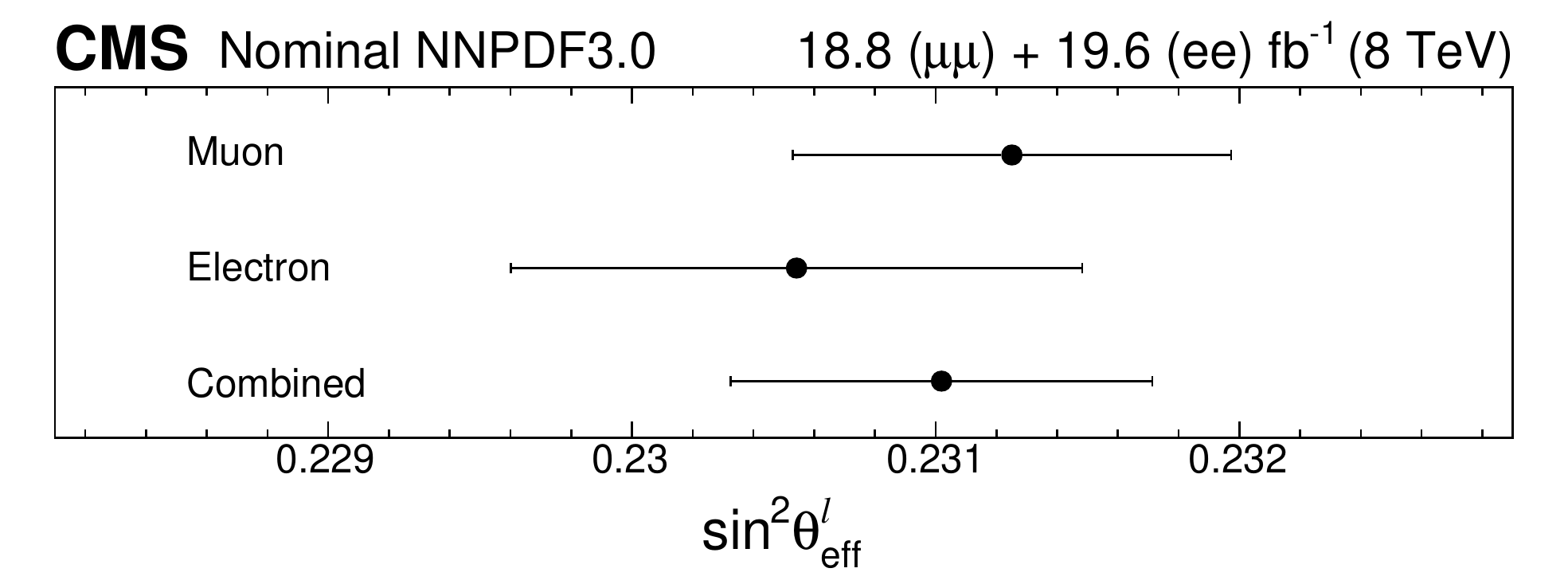}
    \includegraphics[width=0.48\textwidth]{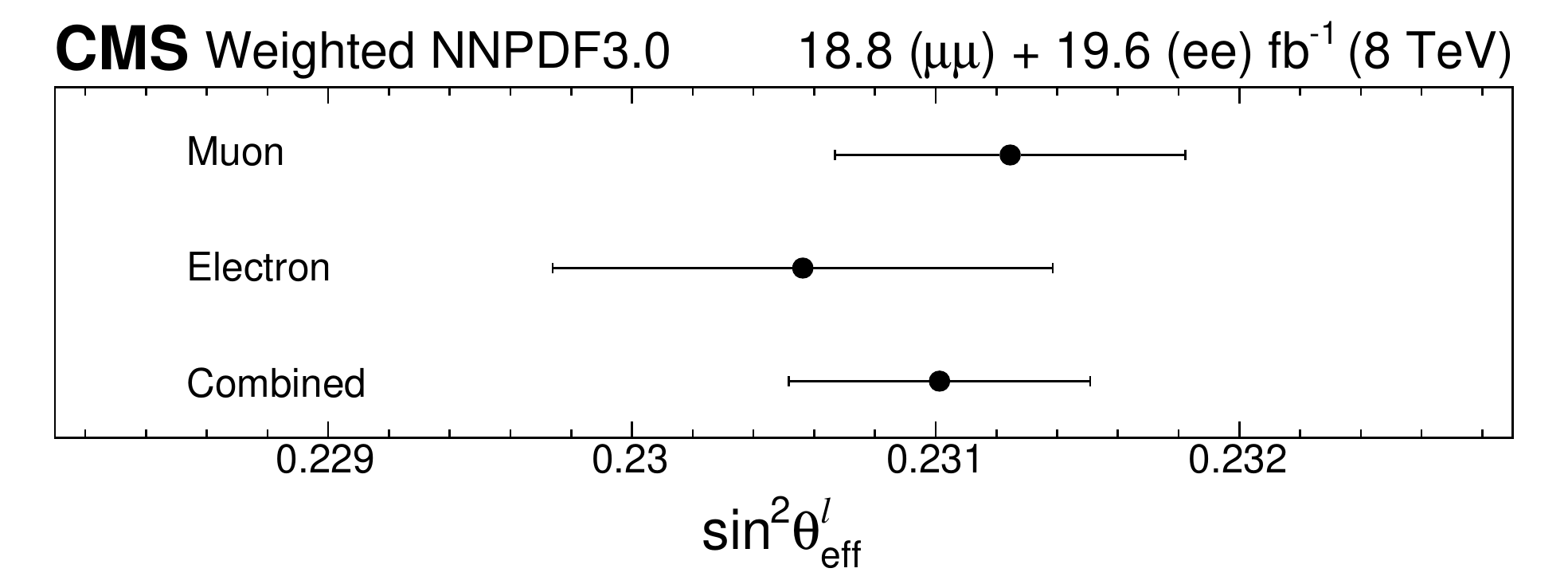}
    \caption{
	The extracted values of \sineff in the muon and electron channels, and their combination.
	The horizontal bars include statistical, experimental, and PDF uncertainties.
	The PDF uncertainties are obtained both without (\cmsLeft) and with (\cmsRight) using the Bayesian $\chi^2$ weighting.
	\label{figure:comination:result}
    }
\end{figure}

Figure~\ref{figure:comination:result} shows the extracted \sineff in the muon and electron decay channels and their combination, with and without constraining the uncertainties in the PDFs. The corresponding numerical values are also listed in Table~\ref{table:combination}. After Bayesian $\chi^2$ reweighting, the PDF uncertainties are  reduced by about a factor of 2. It should be noted that the Bayesian $\chi^2$ reweighting technique works well when the replicas span the optimal value on both of its sides.  In addition, the effective number of replicas after $\chi^2$ reweighting, $n_\txteff=N^2/\sum_{i=1}^{N}w_i^2$, should also be large enough to give a reasonable estimate of the average value and its standard deviation. There are 39 effective replicas after the $\chi^2$ reweighting ($n_\txteff=39$). Including the corresponding statistical uncertainty of 0.00005, the total PDF uncertainty becomes 0.00031. As a cross-check, we perform the analysis with the corresponding set of 1000 NNPDF3.0 replicas in the dimuon channel, and find good consistency between the two results.

\begin{figure}[!htbp]
\centering
    \includegraphics[width=0.48\textwidth]{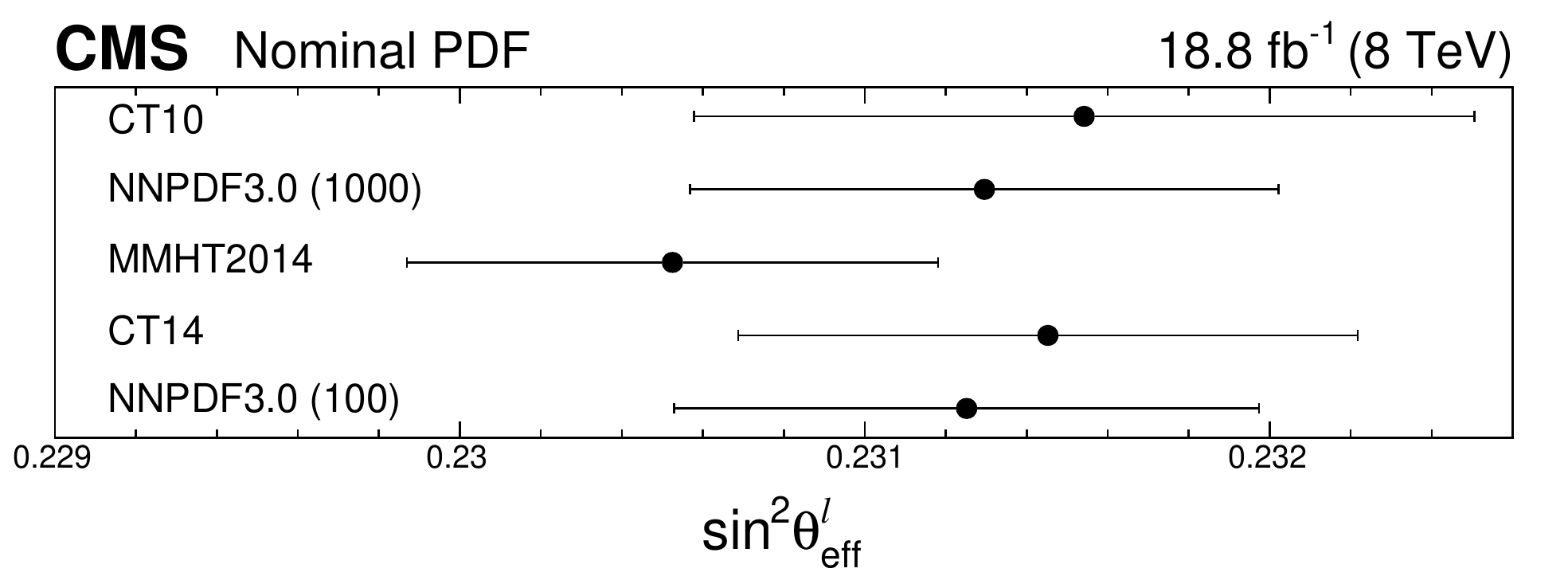}
    \includegraphics[width=0.48\textwidth]{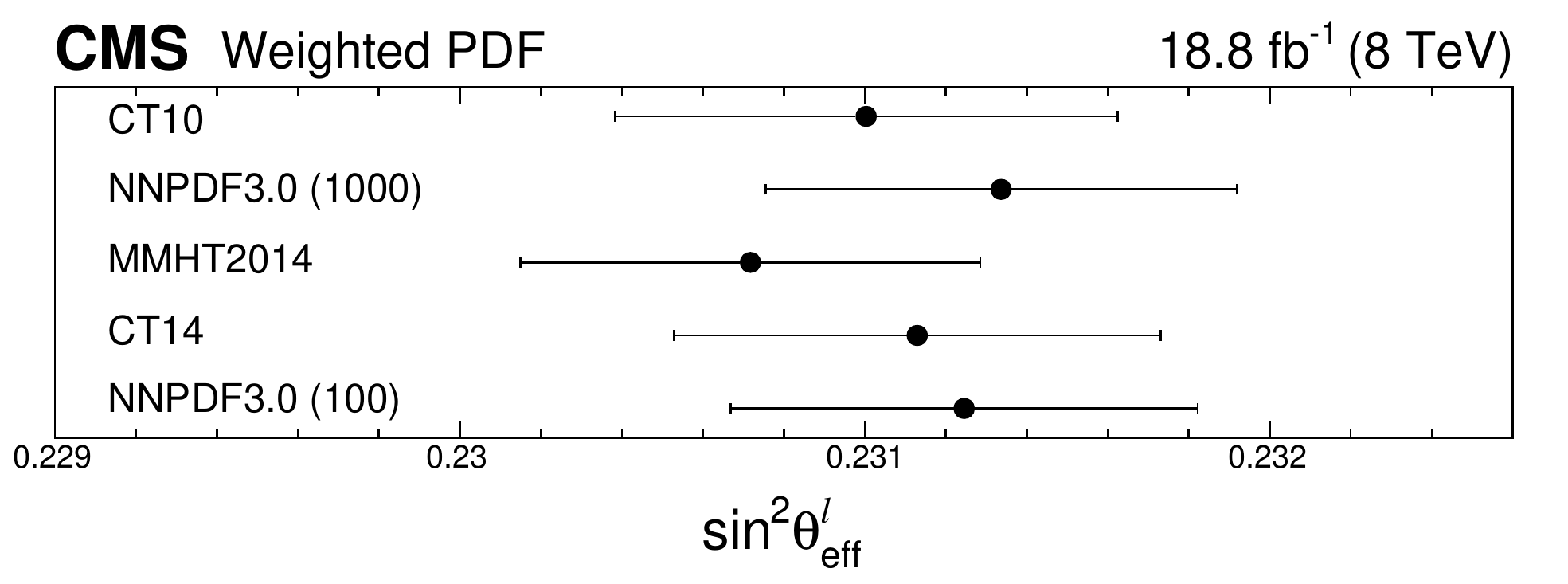}
    \caption{
	Extracted values of \sineff from the dimuon data for different sets of PDFs
	with the nominal (\cmsLeft) and   $\chi^2$-reweighted (\cmsRight) replicas.
	The horizontal error bars include contributions from statistical, experimental, and PDF uncertainties.
	\label{figure:pdffit}
    }
\end{figure}

We have also studied the PDFs represented by Hessian eigenvectors using the CT10~\cite{CTEQ:1007}, CT14~\cite{CT14}, and MMHT2014~\cite{MMHT2014} PDFs in an analysis performed in the dimuon channel. First, we generate the replica predictions ($i$) for each observable $O$ for the Hessian eigensets ($k$):
\begin{equation}
    O_i = O_0 + \frac{1}{2} \sum_{k=0}^{n} (O_{2k+1}-O_{2k+2})R_{ik},
\end{equation}
where $n$ is the number of eigenvector axes, and the $R_{ik}$ are random numbers sampled from the normal distribution with a mean of 0 and a standard deviation of unity. Then, the same technique is applied as used in the NNPDF analysis. The results of fits for these PDFs are summarized in Fig.~\ref{figure:pdffit}. After Bayesian $\chi^2$ reweighting the central predictions for all PDFs are closer to each other, and the corresponding uncertainties are significantly reduced.
The result using CT14 is within about 1/3 of the PDF uncertainty of the NNPDF3.0 result in the muon channel, whereas the MMHT2014 set yields a smaller \sineff value by about one standard deviation. Some of these differences can be reduced by adding more data (e.g. including the electron channel, which is not considered in this check). Some can be attributed to the residual differences in the valence and sea quark distributions, which are not fully constrained using the \AFB distributions alone.  For example, we find that the NLO NNPDF3.0 PDF set yields a very good description for the published 8 TeV CMS muon charge asymmetry ($\chi^2$ of 4.6 for 11 dof).  In contrast, the $\chi^2$ values with the CT14 and MMHT2014 PDF sets are 21.3 and 21.4, respectively. We also constructed a combined set from same number of replicas of NNPDF3.0, CT14, and MMHT2014 PDFs, and after including the data from the \PW\ charge asymmetry in the PDF reweighting, we find the combined weighted average in the dimuon channel differs from the NNPDF3.0 result by only 0.00009, and the standard deviation only increases from 0.00032 to 0.00036.  Consequently, for our quoted results we use only the NNPDF3.0 PDF set, which is used in both dimuon and dielectron analyses.

As an additional test, for the case of Hessian PDFs (including the Hessian NNPDF3.0~\cite{NNPDF30hes})  we perform a simultaneous $\chi^2$ fit for \sineff and all PDF nuisance parameters representing the variations for each eigenvector. As expected for Gaussian distributions, we obtain the same central values and the total uncertainties that are extracted from Bayesian reweighting of the corresponding set of replicas.
\begin{table}[!htbp]
\centering
\topcaption{
    The central value and the PDF uncertainty in the measured \sineff in the muon and electron channels, and their combination,
    obtained without and with constraining PDFs using Bayesian $\chi^2$ reweighting.
}
\label{table:combination}
\begin{tabular}{ l  c  c }
Channel		   &  Not constraining PDFs  & Constraining PDFs \\ \hline
Muons              &  $0.23125\pm0.00054$ & $0.23125\pm0.00032$  \\
Electrons          &  $0.23054\pm0.00064$ & $0.23056\pm0.00045$  \\ [\cmsTabSkip]
Combined           &  $0.23102\pm0.00057$ & $0.23101\pm0.00030$  \\
\end{tabular}
\end{table}

Finally, as a cross-check, we also repeat the measurement using different mass windows for extracting \sineff, and for constraining the PDFs.
Specifically, we first use the central five bins, corresponding to the dimuon mass range of $84<m_{\mu\mu}<95\GeV$, to extract \sineff.
Then, we use predictions based on the extracted \sineff in the lower three $(60 < m_{\mu\mu} <84\GeV)$ and the higher four $(95<m_{\mu\mu}<120\GeV)$ dimuon mass bins, to constrain the PDFs. We find that the statistical uncertainty increases by only about 10\%, and the PDF uncertainty increases by only about 6\% relative to the uncertainties obtained when using the full mass range to extract the \sineff and simultaneously constrain the PDFs. The test thereby confirms that the PDF uncertainties are constrained mainly by the high- and low-mass bins, and that we obtain consistent results with these two approaches.

\section{Summary}

The effective leptonic mixing angle, \sineff, has been extracted  from  measurements of the mass and rapidity dependence of the forward-backward asymmetries \AFB in Drell--Yan $\mu\mu$ and $\Pe\Pe$ production.
As a baseline model, we use the \POWHEG event generator for the inclusive $\Pp\Pp\to\PZ/\gamma\to\ell\ell$ process at leading electroweak order, where the weak mixing angle is interpreted through the improved Born approximation as the effective angle incorporating higher-order corrections.
With more data and new analysis techniques, including precise lepton-momentum calibration, angular event weighting, and additional constraints on PDFs,
the statistical and systematic uncertainties are significantly reduced relative to previous CMS measurements.
The combined result from the dielectron and dimuon channels is:
\ifthenelse{\boolean{cms@external}}{
\begin{multline}
	\sineff=0.23101 \pm 0.00036\stat \pm 0.00018\syst\\
 \pm 0.00016\thy \pm 0.00031\,(\text{PDF}),
\end{multline}

}{
\begin{equation}
	\sineff=0.23101 \pm 0.00036\stat \pm 0.00018\syst \pm 0.00016\thy \pm 0.00031\,(\text{PDF}),
\end{equation}
}
or summing the uncertainties in quadrature,
\begin{equation}
	\sineff=0.23101\pm0.00053.
\end{equation}

\begin{figure*}[!htbp]
\centering
    \includegraphics[width=0.8\textwidth]{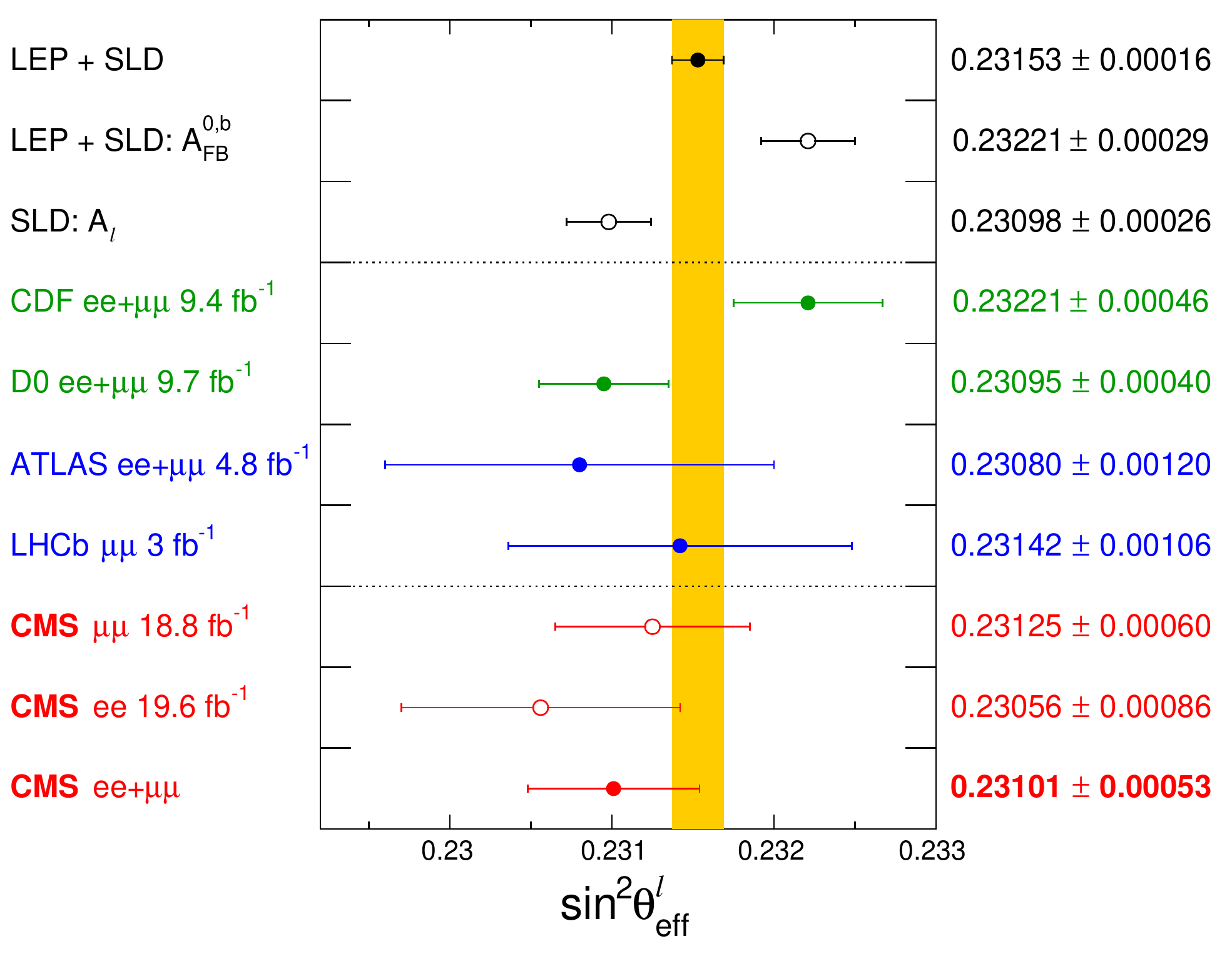}
    \caption{
	Comparison of the measured \sineff in the muon and electron channels and their combination, with
	previous LEP, SLD, Tevatron, and LHC measurements.
	The shaded band corresponds to the combination of the LEP and SLD measurements.
	\label{figure:result}
    }
\end{figure*}

A comparison of the extracted \sineff with previous results from LEP, SLC, Tevatron, and LHC, shown in Fig.~\ref{figure:result},
indicates consistency with the mean of the most precise LEP and SLD results, as well as with the other measurements.

\begin{acknowledgments}

We congratulate our colleagues in the CERN accelerator departments for the excellent performance of the LHC and thank the technical and administrative staffs at CERN and at other CMS institutes for their contributions to the success of the CMS effort. In addition, we gratefully acknowledge the computing centers and personnel of the Worldwide LHC Computing Grid for delivering so effectively the computing infrastructure essential to our analyses. Finally, we acknowledge the enduring support for the construction and operation of the LHC and the CMS detector provided by the following funding agencies: BMWFW and FWF (Austria); FNRS and FWO (Belgium); CNPq, CAPES, FAPERJ, and FAPESP (Brazil); MES (Bulgaria); CERN; CAS, MoST, and NSFC (China); COLCIENCIAS (Colombia); MSES and CSF (Croatia); RPF (Cyprus); SENESCYT (Ecuador); MoER, ERC IUT, and ERDF (Estonia); Academy of Finland, MEC, and HIP (Finland); CEA and CNRS/IN2P3 (France); BMBF, DFG, and HGF (Germany); GSRT (Greece); OTKA and NIH (Hungary); DAE and DST (India); IPM (Iran); SFI (Ireland); INFN (Italy); MSIP and NRF (Republic of Korea); LAS (Lithuania); MOE and UM (Malaysia); BUAP, CINVESTAV, CONACYT, LNS, SEP, and UASLP-FAI (Mexico); MBIE (New Zealand); PAEC (Pakistan); MSHE and NSC (Poland); FCT (Portugal); JINR (Dubna); MON, RosAtom, RAS, RFBR and RAEP (Russia); MESTD (Serbia); SEIDI, CPAN, PCTI and FEDER (Spain); Swiss Funding Agencies (Switzerland); MST (Taipei); ThEPCenter, IPST, STAR, and NSTDA (Thailand); TUBITAK and TAEK (Turkey); NASU and SFFR (Ukraine); STFC (United Kingdom); DOE and NSF (USA).

\hyphenation{Rachada-pisek} Individuals have received support from the Marie-Curie program and the European Research Council and Horizon 2020 Grant, contract No. 675440 (European Union); the Leventis Foundation; the A. P. Sloan Foundation; the Alexander von Humboldt Foundation; the Belgian Federal Science Policy Office; the Fonds pour la Formation \`a la Recherche dans l'Industrie et dans l'Agriculture (FRIA-Belgium); the Agentschap voor Innovatie door Wetenschap en Technologie (IWT-Belgium); the F.R.S.-FNRS and FWO (Belgium) under the "Excellence of Science - EOS" - be.h project n. 30820817; the Ministry of Education, Youth and Sports (MEYS) of the Czech Republic; the Council of Science and Industrial Research, India; the HOMING PLUS program of the Foundation for Polish Science, cofinanced from European Union, Regional Development Fund, the Mobility Plus program of the Ministry of Science and Higher Education, the National Science Center (Poland), contracts Harmonia 2014/14/M/ST2/00428, Opus 2014/13/B/ST2/02543, 2014/15/B/ST2/03998, and 2015/19/B/ST2/02861, Sonata-bis 2012/07/E/ST2/01406; the National Priorities Research Program by Qatar National Research Fund; the Programa Severo Ochoa del Principado de Asturias; the Thalis and Aristeia programs cofinanced by EU-ESF and the Greek NSRF; the Rachadapisek Sompot Fund for Postdoctoral Fellowship, Chulalongkorn University and the Chulalongkorn Academic into Its 2nd Century Project Advancement Project (Thailand); the Welch Foundation, contract C-1845; and the Weston Havens Foundation (USA).

\end{acknowledgments}

\bibliography{auto_generated}

\cleardoublepage \appendix\section{The CMS Collaboration \label{app:collab}}\begin{sloppypar}\hyphenpenalty=5000\widowpenalty=500\clubpenalty=5000\vskip\cmsinstskip
\textbf{Yerevan Physics Institute, Yerevan, Armenia}\\*[0pt]
A.M.~Sirunyan, A.~Tumasyan
\vskip\cmsinstskip
\textbf{Institut f\"{u}r Hochenergiephysik, Wien, Austria}\\*[0pt]
W.~Adam, F.~Ambrogi, E.~Asilar, T.~Bergauer, J.~Brandstetter, E.~Brondolin, M.~Dragicevic, J.~Er\"{o}, A.~Escalante~Del~Valle, M.~Flechl, R.~Fr\"{u}hwirth\cmsAuthorMark{1}, V.M.~Ghete, J.~Hrubec, M.~Jeitler\cmsAuthorMark{1}, N.~Krammer, I.~Kr\"{a}tschmer, D.~Liko, T.~Madlener, I.~Mikulec, N.~Rad, H.~Rohringer, J.~Schieck\cmsAuthorMark{1}, R.~Sch\"{o}fbeck, M.~Spanring, D.~Spitzbart, A.~Taurok, W.~Waltenberger, J.~Wittmann, C.-E.~Wulz\cmsAuthorMark{1}, M.~Zarucki
% ...
\vskip\cmsinstskip
% ...
1:  Also at Vienna University of Technology, Vienna, Austria\\
% ...
\end{sloppypar}
\end{document}